\documentclass[sigconf]{acmart}

\AtBeginDocument{%
  }

\usepackage{float}
\usepackage{amsmath,amsfonts}
\usepackage{nicefrac}
\usepackage{multirow}
\usepackage{algorithm}
\usepackage{algorithmicx}
\usepackage{algpseudocode}
\usepackage{tikz}
\usepackage{pgfplots}
\pgfplotsset{compat=1.18}
\usetikzlibrary{shapes,arrows,positioning,fit,backgrounds}
\usepackage{listings}
\usepackage{array}
\usepackage{ragged2e}
\usepackage{wrapfig}
\usepackage{tabularx}

\definecolor{codegreen}{rgb}{0,0.6,0}
\definecolor{codegray}{rgb}{0.5,0.5,0.5}
\definecolor{codepink}{RGB}{252, 142, 172}
\definecolor{codepurple}{rgb}{0.58,0,0.82}
\definecolor{backcolour}{RGB}{245,245,245}

\lstdefinestyle{promptstyle}{
    backgroundcolor=\color{backcolour},
    commentstyle=\color{magenta},
    keywordstyle=\color{blue},
    numberstyle=\tiny\color{codegray},
    stringstyle=\color{codepurple},
    basicstyle=\fontfamily{\ttdefault}\footnotesize,
    breakatwhitespace=false,
    breaklines=true,
    keepspaces=true,
    frame=single,
    numbersep=5pt,
    showspaces=false,
    showstringspaces=false,
    showtabs=false,
    tabsize=2,
    classoffset=1,
    keywordstyle=\color{violet},
    classoffset=0,
}

\newcolumntype{L}[1]{>{\raggedright\arraybackslash}p{#1}}
\providecommand{\checkmark}{\ding{51}}

\copyrightyear{2027}
\acmYear{2027}
\setcopyright{acmlicensed}
\acmConference[KDD '27]{Proceedings of the 33rd ACM SIGKDD Conference on Knowledge Discovery and Data Mining V.1}{August 1--5, 2027}{San Jose, United States}
\acmBooktitle{Proceedings of the 33rd ACM SIGKDD Conference on Knowledge Discovery and Data Mining V.1 (KDD '27), August 1--5, 2027, San Jose, United States}
\acmDOI{}
\acmISBN{}

\begin{document}

\title{AutoDFT: A Closed-Loop Multi-Agent Framework for Autonomous DFT Calculations}

\author{Penghui Yang}
\authornote{These authors contributed equally to this research.}
\affiliation{%
  \institution{College of Computing and Data Science, Nanyang Technological University}
  \country{Singapore}
}

\author{Zhonghan Zhang}
\authornotemark[1]
\affiliation{%
  \institution{College of Computing and Data Science, Nanyang Technological University}
  \country{Singapore}
}

\author{Yue Li}
\authornotemark[1]
\affiliation{%
  \institution{School of Materials Science and Engineering, Nanyang Technological University}
  \country{Singapore}
}

\author{Xinrun Wang}
\authornote{Corresponding authors.}
\affiliation{%
  \institution{School of Computing and Information Systems, Singapore Management University}
  \country{Singapore}
}
\email{xrwang@smu.edu.sg}

\author{Yanchen Deng}
\affiliation{%
  \institution{College of Computing and Data Science, Nanyang Technological University}
  \country{Singapore}
}

\author{Yuhao Lu}
\affiliation{%
  \institution{College of Computing and Data Science, Nanyang Technological University}
  \country{Singapore}
}

\author{Bijun Tang}
\authornotemark[2]
\affiliation{%
  \institution{School of Materials Science and Engineering, Nanyang Technological University}
  \country{Singapore}
}
\email{bjtang@ntu.edu.sg}

\author{Zheng Liu}
\affiliation{%
  \institution{School of Materials Science and Engineering, Nanyang Technological University}
  \country{Singapore}
}

\author{Bo An}
\authornotemark[2]
\affiliation{%
  \institution{College of Computing and Data Science, Nanyang Technological University}
  \country{Singapore}
}
\email{boan@ntu.edu.sg}

\renewcommand{\shortauthors}{Penghui Yang, Zhonghan Zhang, Yue Li, Xinrun Wang, Yanchen Deng, Yuhao Lu, Bijun Tang, Zheng Liu, Bo An}

\begin{abstract}
Density functional theory (DFT) serves as the basis for computational discovery in materials science and chemistry, yet each calculation demands extensive human effort: adjusting algorithms when convergence stalls, revising plans when unexpected physics emerges, and inserting steps as intermediate results reshape the problem. Existing LLM-based agents automate only the initial planning stage, producing a full execution plan upfront and leaving all subsequent adaptation to hand-crafted rules. As a result, these workflows remain fragile, do not generalize well beyond pre-planned scenarios, and often require expert intervention when failures or unexpected intermediate results require changes to the calculation path. Here, we introduce \textbf{AutoDFT}, a closed-loop multi-agent framework that embeds LLM reasoning into every stage of the DFT lifecycle, where a strategic planner produces a skeletal plan of step objectives; a step planner generates numerical parameters just in time from preceding results; and a monitor-recover-reflect cycle diagnoses failures, repairs them, and revises the plan when the evidence justifies it. We demonstrate both breadth and depth: breadth on \textbf{VASPBench}, a purpose-built benchmark spanning 34 tasks and 9 DFT calculation types, where AutoDFT achieves 94.1\% task-level success with GPT-5.2; and depth on established materials databases, where AutoDFT produces quantitatively reliable property predictions across electronic, magnetic, and energetic properties. By closing the loop between planning and execution, AutoDFT enables experimentalists without deep computational expertise to obtain reliable first-principles results. Our code is available at: \url{https://github.com/penghui-yang/AutoDFT}.
\end{abstract}



\maketitle

\begin{figure*}[t]
    \centering
    \includegraphics[width=0.85\linewidth]{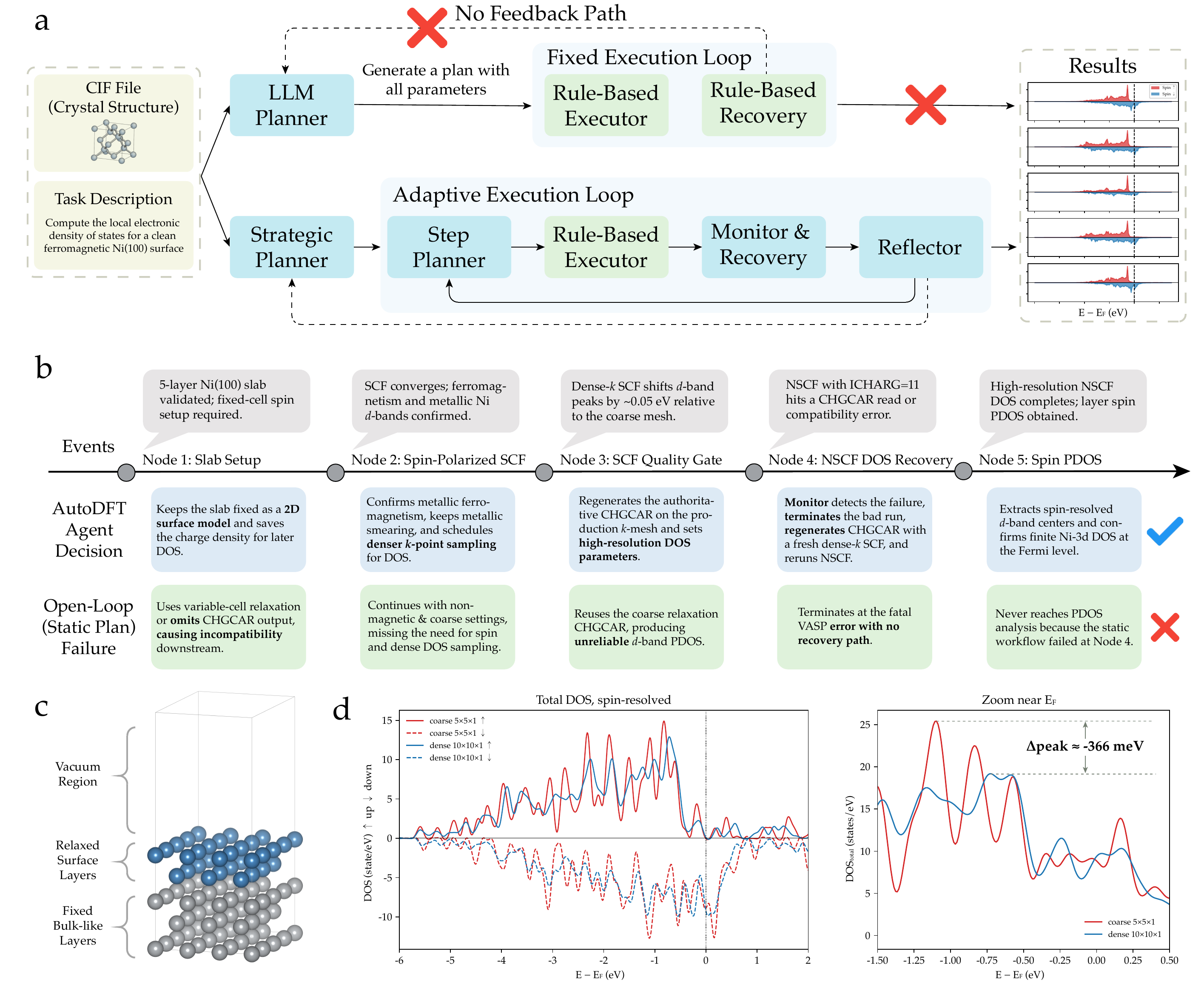}
    \caption{Open-loop vs. closed-loop DFT automation. (a) Comparison of existing open-loop approaches, which fix all parameters at planning time and rely on rule-based execution, with AutoDFT's closed-loop architecture, which combines hierarchical planning decomposition with LLM-driven monitoring, recovery, and plan revision throughout the calculation. (b) A layer-resolved spin Projected Density of States (PDOS) calculation for a Ni(100) slab illustrating the practical impact of closed-loop execution. A static plan may relax the slab as a bulk-like 3D cell, miss emergent ferromagnetism, reuse an incompatible or under-converged charge density, and fail before reaching the target PDOS. However, AutoDFT can adapt at each stage, ultimately producing a converged layer-resolved spin PDOS. (c) Structure visualization of the five-layer Ni(100) slab. (d) Spin-resolved DOS comparison between coarse and dense $k$-point meshes, showing a near-Fermi $d$-band peak shift that motivates $k$-point refinement before the final PDOS calculation. All VASP-related acronyms, file names, and input tags are expanded in Appendix~\ref{app:glossary}.}
    \label{fig:teaser}
\end{figure*}

\section{Introduction}

Density functional theory (DFT) has been instrumental in predicting properties of materials and molecules from first principles, enabling breakthroughs in areas ranging from energy storage~\citep{urban2016computational} and semiconductors~\citep{curtarolo2013high} to catalysis~\citep{norskov2009towards} and low-dimensional materials~\citep{mounet2018two}. 
Despite its central role, practical DFT workflows remain heavily manual. A typical study consists of a sequence of interdependent computational steps, each requiring researchers to choose exchange-correlation functionals, configure numerical parameters (plane-wave cutoff, $k$-point density, smearing width, etc.), manage convergence criteria, diagnose failures, and interpret intermediate outputs before advancing to the next computational step.
This expertise bottleneck prevents truly autonomous computational campaigns, making end-to-end automation difficult.

Recent advances in large language models (LLMs) have demonstrated strong capabilities in scientific reasoning~\citep{boiko2023autonomous}, code generation~\citep{li2022competition}, and autonomous agent design~\citep{yao2023react}. Within computational materials science, LLMs have been applied to property prediction~\citep{niyongabo2025llm}, literature mining~\citep{foppiano2024mining, huang2026matseek}, and preliminary workflow generation~\citep{ghafarollahi2025atomagents}. However, these efforts automate only the initial planning stage, producing a complete execution plan upfront and delegating all subsequent adaptation to hand-crafted rules. It remains an open challenge to achieve fully autonomous, closed-loop execution of first-principles calculations in which agents not only design high-level workflows but also monitor progress, recover from failures, and revise strategies grounded in physical reasoning.

Several factors make this problem particularly challenging. First, even a physically straightforward task, such as bandgap determination, may require a multi-step pipeline (relaxation $\rightarrow$ static self-consistent field (SCF) calculation $\rightarrow$ band structure), where each step involves a large number of input parameters whose optimal values depend on prior results. Second, DFT calculations fail for various physical and numerical reasons, including charge sloshing, unconverged self-consistent cycles, symmetry breaking, and inadequate $k$-point sampling, and each failure mode demands a distinct recovery strategy that cannot be anticipated at planning time. Third, the optimal workflow for a given material is often unknown a priori; emergent properties, such as unexpected magnetism or topological character, discovered only during calculation, may necessitate extending or restructuring the plan. Figure~\ref{fig:teaser} provides an overview of the open-loop failure modes addressed by AutoDFT and illustrates how closed-loop execution resolves them in a Ni(100) surface PDOS workflow.

We address these challenges with AutoDFT, a closed-loop seven-agent framework built on top of VASP\footnote{VASP (Vienna Ab initio Simulation Package)~\citep{kresse1996vasp} is the most widely used plane-wave DFT code. Its foundational paper ranks among the ten most-cited scientific articles of all time~\citep{van2025these}. Our framework is demonstrated based on VASP, and is adaptable for other DFT software packages, e.g., Quantum ESPRESSO and Gaussian.} that embeds LLM reasoning into every stage of the DFT lifecycle. The design rests on two core principles: (i) hierarchical planning decomposition, which separates strategic workflow design from tactical parameter generation; and (ii) closed-loop adaptive execution, in which real-time monitoring, LLM-driven failure recovery, and evidence-based plan revision allow the system to respond to emergent computational and physical findings throughout the calculation. To implement these principles, AutoDFT assigns complementary responsibilities to seven agents: a strategic planner that produces a skeletal plan of step-level objectives; a step planner that generates concrete numerical parameters just in time from preceding results; and a monitor–recover–reflect cycle that runs the VASP job, diagnoses runtime failures, performs repairs, and revises the downstream plans.

Our contributions are threefold. First, we propose \textbf{AutoDFT}, a seven-agent architecture for end-to-end DFT workflow automation, whose modular design separates planning, parameter generation, parameter translation, monitoring, recovery, and reflection while supporting 30 material properties across a broad range of common DFT calculation types. Second, we introduce \textbf{VASPBench}, a curated benchmark comprising 34 DFT calculation tasks with ground-truth outputs derived from official VASP documentation, spanning 9 distinct calculation types. Third, we conduct extensive experiments on both VASPBench and the Materials Project database, together with comparisons against rule-based and open-loop baselines, demonstrating that AutoDFT achieves 85.3–94.1\% task-level success rates and that closed-loop adaptation provides consistent gains beyond initial LLM planning.
\section{Related Work}

\textbf{Workflow infrastructure for computational materials science.}
Systems such as AiiDA~\citep{pizzi2016aiida}, FireWorks~\citep{jain2015fireworks}, and Atomate2~\citep{ganose2025atomate2} provide programmatic orchestration of first-principles calculations, and high-throughput frameworks including the Materials Project~\citep{jain2013commentary} and AFLOW~\citep{curtarolo2012aflow} have enabled property screening at scale. These platforms, however, rely on expert-authored, fixed workflows: each supported calculation type requires manual specification of input parameters, convergence criteria, and failure-handling rules, leaving no mechanism for adaptive reasoning when confronted with novel materials or unexpected computational behavior.

\textbf{LLM-assisted DFT automation.}
Recent work has begun to apply LLMs to automate DFT workflow generation. VASPilot~\citep{liu2025vaspilot} and AMLP~\citep{lahouari2025automated} employ multi-agent architectures to translate task descriptions into VASP input files, automating the planning stage of the computation. In both systems, simulation parameters are derived from library defaults rather than from LLM reasoning conditioned on the material and prior results, and the workflow structure remains fixed throughout execution. DREAMS~\citep{wang2025dreams} extends LLM participation to error recovery through a dedicated convergence agent that analyzes input and output files via LLM reasoning to suggest parameter adjustments for failed calculations; its supervisor agent can also insert missing computational steps during execution. However, the adaptive behavior of DREAMS operates at the level of numerical convergence: when a calculation fails to converge, the convergence agent diagnoses the issue and proposes solver-parameter adjustments. The system does not re-evaluate the scientific strategy when a step succeeds with unexpected results. These systems demonstrate the value of LLMs for DFT automation while leaving workflow-level adaptation outside the scope of learned reasoning.

\textbf{LLM agents for broader scientific automation.}
Beyond computational materials science, LLM-based agents have been applied to organic synthesis planning~\citep{boiko2023autonomous}, protein structure prediction orchestration~\citep{xu2026proteinmcp}, and autonomous experiment design~\citep{zhou2025autonomous}. AtomAgents~\citep{ghafarollahi2025atomagents} employs a multi-agent architecture for alloy design but operates over static workflow templates, and ChemCrow~\citep{bran2024chemcrow} integrates chemical tools without addressing first-principles calculations. While these systems illustrate the breadth of LLM-driven scientific automation, none of them target the full lifecycle of a DFT computation, from adaptive planning to failure recovery, with closed-loop LLM reasoning.

\textbf{Agentic reasoning and tool use.}
General-purpose agent frameworks such as ReAct~\citep{yao2023react}, Reflexion~\citep{shinn2024reflexion}, and AutoGen~\citep{wu2024autogen} have established patterns for interleaving LLM reasoning with tool invocation. When applied directly to DFT automation, these patterns encounter several structural mismatches: VASP parameters are highly interdependent, making per-parameter trial and error combinatorially expensive; individual calculations can run for hours, violating the rapid-feedback assumption underlying ReAct-style loops; and a numerically converged calculation can still be physically wrong, which cannot be captured by generic success/failure feedback. AutoDFT addresses these challenges through a hierarchical architecture where a step planner generates internally consistent parameters in a single call, a dual-path monitor amortizes LLM cost over long-running jobs, and a step reflector performs domain-specific plausibility checks after each successful step.
\section{AutoDFT}
\label{sec:method}

\begin{figure*}[t]
    \centering
    \includegraphics[width=0.8\linewidth]{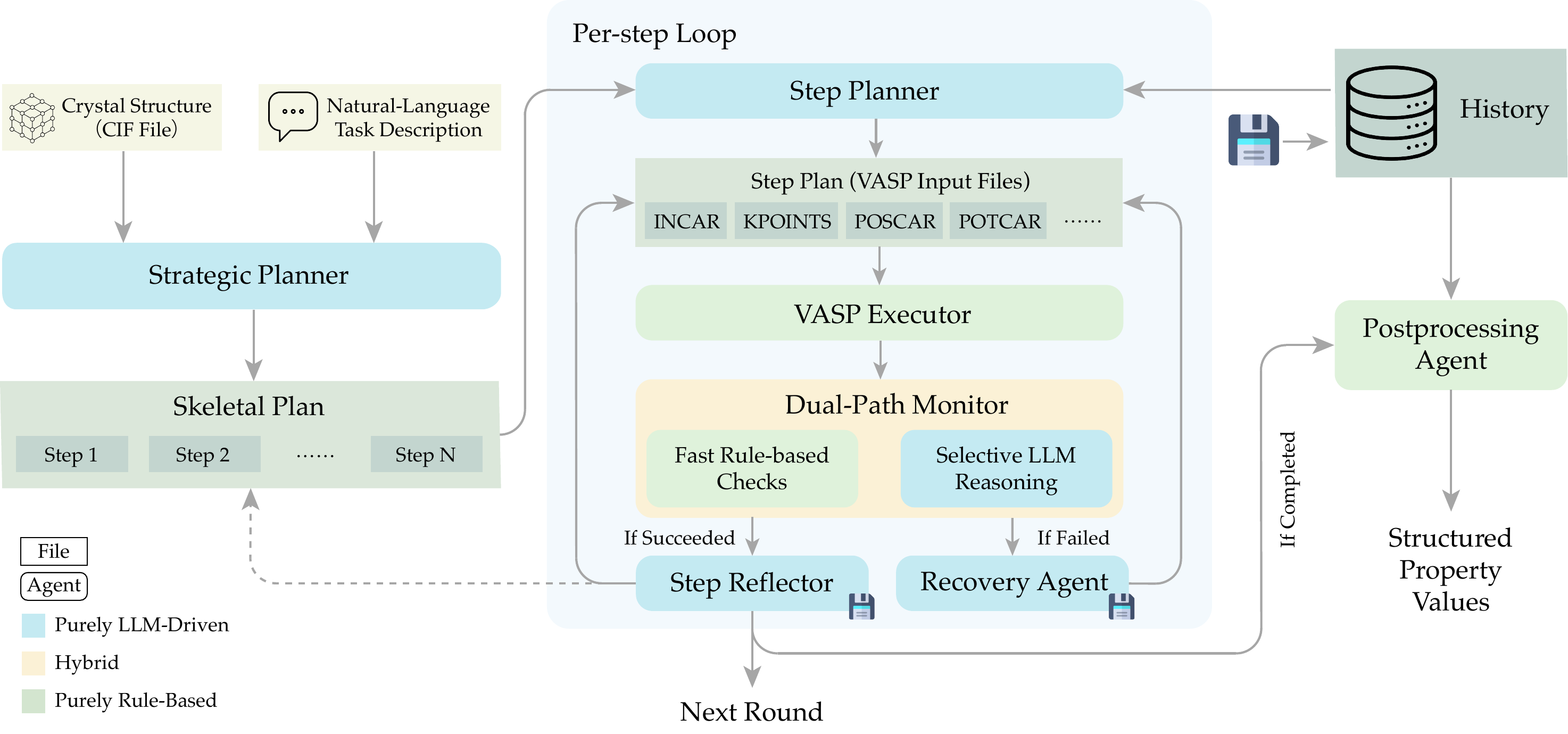}
    \caption{Overview of the AutoDFT framework. Given a crystal structure and task description, the Strategic Planner generates a skeletal plan (step objectives without concrete parameters). Execution then enters a per-step loop: the Step Planner materializes standard and task-dependent VASP input files using the accumulated History, the VASP Executor submits the job, and the Dual-Path Monitor tracks convergence. Failed steps enter an inner recovery loop; successful steps are evaluated by the Step Reflector, which may accept the result, redo the step, or revise the skeletal plan (dashed path back to the Strategic Planner). The History store records completed-step inputs, monitor/recovery decisions, convergence traces, and outcome summaries, providing accumulated context for step planning, reflection, and final postprocessing. Module background colors indicate the reasoning paradigm: blue denotes LLM-driven components, pale yellow denotes hybrid components with selective LLM reasoning, and green denotes purely rule-based components.}
    \label{fig:architecture}
\end{figure*}


A typical DFT study proceeds through a sequence of dependent computation steps, where each step both depends on the outputs of its predecessors and may fail in ways that require expert-level diagnosis. Existing LLM-driven frameworks treat this process as a two-phase pipeline: an LLM generates a complete workflow plan upfront, and a rule-based engine executes it without further reasoning. This \emph{open-loop} design collapses when a step produces unexpected results (e.g., a relaxation that breaks symmetry, or an SCF that converges to a metallic state for a nominally semiconducting material), because the downstream plan was committed before these outcomes were observed.

AutoDFT replaces this open-loop pipeline with a \emph{closed-loop} architecture organized around two design principles. The first, \emph{hierarchical planning decomposition}, separates strategic decisions (what sequence of computations to run) from tactical decisions (which VASP parameters to set for each computation). The second, \emph{closed-loop adaptive execution}, ensures that every computation step is followed by LLM-driven evaluation, so that failures trigger reasoned recovery and successful but unexpected results can trigger mid-workflow re-planning.

Figure~\ref{fig:architecture} illustrates the resulting system. Given a crystal structure (CIF file) and a natural-language task description, the \textbf{Strategic Planner} generates a skeletal plan: an ordered sequence of high-level steps with objectives and success criteria, but without concrete VASP parameters. AutoDFT maintains a workflow-level \textbf{History} store that records completed-step inputs, monitor verdicts, recovery attempts, convergence traces, and \texttt{StepOutcomeSummary} records. Execution then enters a per-step loop. For each step, the \textbf{Step Planner} materializes VASP parameters informed by this accumulated History. The \textbf{VASP Executor} prepares input files and submits the job. While the job runs, the \textbf{Dual-Path Monitor} combines fast rule-based checks with selective LLM reasoning to detect convergence issues. If a step fails, the \textbf{Recovery Agent} diagnoses the root cause and generates modified parameters for a retry. If a step succeeds, the \textbf{Step Reflector} evaluates the physical plausibility of the output (for example, checking whether a relaxed structure has preserved its expected symmetry or whether the computed energy falls within a reasonable range) and decides whether to accept the result, redo the step with more stringent settings, or revise the skeletal plan for remaining steps. Upon completion of all steps, a rule-based \textbf{Postprocessing Agent} reads the final History and associated VASP output files to extract structured property values. The seven agents span a spectrum from purely LLM-driven (Strategic Planner, Step Planner, Recovery Agent, Step Reflector) through hybrid (Monitor) to purely rule-based (Executor, Postprocessing Agent), allocating LLM reasoning where scientific judgment is required and avoiding it where deterministic logic suffices.

\subsection{Hierarchical Planning}
\label{sec:planning}

Generating a complete DFT workflow requires two qualitatively different kinds of reasoning. Strategic reasoning determines the sequence of computations and their interdependencies: a band-structure calculation requires a prior relaxation and static SCF, spin-orbit coupling demands a preceding collinear magnetic ground state, and so on. Tactical reasoning determines the concrete parameters for each computation, such as cutoff energies, $k$-point densities, and convergence thresholds, and these choices often depend on the outcomes of earlier steps. Conflating the two in a single LLM call, as prior systems do, forces the model to commit to tactical details without the necessary information.

\textbf{Strategic Planner.}
The Strategic Planner receives the crystal structure and the task description. It first converts the structure to VASP input format via \texttt{vaspkit}~\citep{wang2021vaspkit}, then infers material context from the input: dimensionality (2D slab vs.\ 3D bulk) from the lattice geometry, magnetic character from the elemental composition, and so forth. Based on this context, it generates a \texttt{SkeletalPlan} through a single LLM call. Each entry in the plan specifies a step name, a natural-language objective, a computation type (relaxation, static SCF, etc.), dependencies on prior steps, physical constraints, and success criteria, but no concrete simulation parameters.
 
\textbf{Step Planner.}
Before each step executes, the Step Planner materializes its concrete VASP parameters. The planner receives the skeletal step specification together with the \texttt{StepOutcomeSummary} records from all completed steps, which contain energy trajectories, convergence profiles, forces, symmetry changes, and any warnings. A single LLM invocation generates the full set of simulation parameters, including numerical settings that control the calculation (INCAR), the reciprocal-space sampling grid (KPOINTS), and any preprocessing actions. When the Step Reflector has requested a redo with specific parameter modifications, those modifications are applied directly rather than being re-generated. This step-level replanning is the core of adaptivity in AutoDFT. Because the Step Planner observes the full execution history at every invocation, it can adjust parameters in response to emergent behavior from earlier steps, which is impossible in an open-loop system where all parameters are fixed at planning time.

\subsection{Closed-Loop Execution}
\label{sec:execution}

Once the Step Planner produces concrete parameters, execution enters a cycle of run--monitor--recover--reflect that continues until the step succeeds or retry limits are exhausted. The four components described below are tightly coupled: the VASP Executor launches a job whose runtime output is continuously consumed by the Dual-Path Monitor; a non-\textsc{Continue} verdict from the monitor triggers the Recovery Agent, which feeds revised parameters back to the VASP Executor; and upon convergence the Step Reflector decides whether the result is physically sound or whether the current step must be revised before the loop advances.

\textbf{VASP Executor.}
The VASP Executor is a deterministic layer that translates \texttt{StepParameters} into a VASP working directory: it writes the input files that specify simulation parameters (INCAR), atomic coordinates (POSCAR), pseudopotentials (POTCAR), and the reciprocal-space sampling grid (KPOINTS), together with any task-dependent auxiliary files for the current step. It then submits the job, and returns file handles to the monitor. It performs no reasoning and incurs no LLM cost.

\textbf{Dual-Path Monitor.}
Runtime monitoring must balance responsiveness against LLM cost: frequent LLM queries become expensive because a single VASP job may run for over 10 hours, even on an H100 GPU. AutoDFT addresses this with a two-level architecture. A \emph{rule-based monitor} parses the energy-convergence log (OSZICAR) at a fixed check interval of ten minutes, detecting electronic non-convergence, energy oscillation, force stagnation, or error keywords in the detailed output log (OUTCAR). This path handles unambiguous failure modes with zero LLM overhead. A \emph{selective LLM monitor} is invoked only when the rule path returns a non-\textsc{Continue} status or at a periodic fallback interval of every five rule-based checks. In practice, a typical relaxation step triggers about 50 rule-based checks over its runtime but only 1--4 LLM monitor calls, reducing LLM invocations by an order of magnitude compared with calling the LLM at every check. When invoked, the LLM receives the last 30 lines of the convergence log, an output summary, the step parameters, and the physical context, and returns a structured decision (\textsc{Continue}, \textsc{Converged}, or \textsc{Terminate}) together with a reasoning trace. The selective design retains the ability to detect subtle issues that escape rule-based heuristics, such as slow drift toward a metastable magnetic configuration or gradual loss of symmetry during ionic relaxation, without incurring the latency and token cost of per-check LLM calls.

\textbf{Recovery Agent.}
When the monitor issues a \textsc{Terminate} verdict, the Recovery Agent diagnoses the failure and generates modified parameters for a retry. It receives the failed step's parameters, the convergence and output logs, the error classification from the monitor, and the history of any prior recovery attempts for the same step, and reasons over this joint context to propose targeted parameter adjustments. The failed working directory is archived before each retry to preserve the diagnostic record. Retries are capped at three per step; if all attempts are exhausted, control passes to the Step Reflector, which may choose to restructure the remaining plan around the unresolvable failure.

\textbf{Step Reflector.}
After a step converges, the Step Reflector evaluates the physical reasonableness of the result. It receives the \texttt{StepOutcomeSummary} together with the current skeletal plan and material context, and returns one of three decisions. \textsc{Accept} advances to the next planned step. \textsc{Redo} repeats the current step with tightened parameters (e.g., a denser reciprocal-space grid or stricter force threshold) while leaving the plan unchanged. \textsc{Modify} revises the skeletal plan itself; the three admissible operations are \emph{insert} (add a step, e.g., an additional convergence test), \emph{drop} (remove a step that is no longer justified by the observed results), and \emph{replace} (substitute a step with a different calculation type). Each operation must satisfy a dependency invariant: every step in the revised plan must have its prerequisite outputs produced by an earlier step. To prevent unbounded plan growth, plan modifications are capped at $R_{\max}=3$ per workflow and total steps at $N_{\max}=15$; both values were chosen conservatively based on the observation that the longest manually designed VASP workflows in our benchmark rarely exceed seven steps or require more than two mid-course corrections. A detailed re-planning example is provided in the case study (Appendix~\ref{sec:case-studies}).
\subsection{Property Extraction}

Upon completion of all workflow steps, a rule-based Postprocessing Agent extracts structured property values from the VASP output files. Supported properties span electronic (bandgap, band character, density of states), structural (space group, equilibrium volume), magnetic (total and atomic magnetic moments), energetic (total energy, formation energy), and spectroscopic (phonon frequencies, Raman-active modes, dielectric constants) quantities. The complete property coverage is summarized in Appendix~\ref{app:task03-supported-properties}. Because property extraction is deterministic parsing of well-defined file formats, it requires no LLM reasoning and operates at negligible cost.

\section{Experiments}
\label{sec:experiments}

We evaluate AutoDFT along three axes: (i) end-to-end task success on a benchmark of realistic VASP workflows, (ii) physical correctness of the properties extracted from successful runs, and (iii) the sources of improvement provided by closed-loop execution. The experiments are designed to test the central claim of this work: closed-loop adaptive execution provides measurable gains over the open-loop planning paradigm adopted by prior DFT automation systems.

\subsection{Experimental Setup}
\label{sec:setup}

\textbf{VASPBench.}
We construct VASPBench, a benchmark of 34 DFT tasks derived from the official VASP documentation. Each task provides a target structure and a natural-language calculation objective, and the system receives no additional hints about parameter choices or workflow structure. We evaluate the task-level success rate, defined as the fraction of tasks that produce converged outputs satisfying the benchmark validation checks. The full task composition is reported in Appendix~\ref{app:task01-vaspbench}.

\textbf{Materials Project subset.}
To assess physical correctness beyond binary task completion, we evaluate AutoDFT on a 20-material subset from the Materials Project database~\citep{jain2013commentary}. For each material, we compare the electronic bandgap, total magnetic moment, and formation energy against the reference values. We report the mean absolute error (MAE) for each property over successfully completed runs. Details of the subset construction are provided in Appendix~\ref{app:task11-mp20}.

\textbf{Baselines and backbones.}
We compare \textit{AutoDFT-Full}, the complete system described in Section~\ref{sec:method}, with two baselines. \textit{Rule-Based} executes a hand-crafted workflow template selected from the task description without using an LLM. \textit{AutoDFT-OpenLoop} uses one initial LLM call to generate the full workflow in advance, followed by deterministic execution, monitoring, and recovery. We instantiate AutoDFT with GPT-5.2~\citep{openai2025gpt52}, Claude Sonnet 4.6~\citep{anthropic2026sonnet46}, and Gemini 3.1 Pro Preview~\citep{deepmind2026gemini31} to test whether the benefits of closed-loop execution are consistent across model families. Details are in Appendix~\ref{app:imple}.

\textbf{Additional ablation baselines.}
To distinguish the effects of upfront parameter generation and runtime adaptation, we introduce two intermediate configurations. \textit{Static-Plan} generates the complete workflow and all VASP parameters (including INCAR and KPOINTS) in a single initial LLM call, rather than using the heuristic parameter templates of OpenLoop. \textit{Static-Plan + LLM Repair} additionally allows LLM-driven recovery from VASP execution errors, but does not employ history-conditioned per-step planning, LLM-based runtime monitoring, or post-convergence reflection. These configurations provide more controlled comparisons between static planning, reactive error recovery, and the complete closed-loop framework.

\subsection{VASPBench: End-to-End Task Success}
\label{sec:main-results}

\begin{table*}[t]
\centering
\caption{End-to-end success and monetary cost on VASPBench (34 tasks). Tok./task denotes average total LLM tokens per task. GPU cost is estimated at \$2 per H100 GPU-hour, following the cost model in Appendix~\ref{app:execution-cost}. LLM inference cost is estimated from total token usage using rates of \$14/M tokens for GPT-5.2, \$15/M tokens for Claude Sonnet 4.6, and \$12/M tokens for Gemini 3.1 Pro Preview. Cost/task includes both VASP execution and LLM inference. Cost/succ. denotes total cost divided by the number of successful tasks. In this table, ``Sonnet 4.6'' is short for ``Claude Sonnet 4.6'', and ``Gemini 3.1'' is short for ``Gemini 3.1 Pro Preview''. The best success rate and lowest cost/succ. among LLM-based configurations for each backbone are shown in bold.}
\label{tab:main-results}
\begin{tabular}{cccccccc}
\toprule
Backbone & Config. & Succ. & Succ. (\%) & Tok. / Task & GPU-h / Task & Cost / Task & Cost / Succ. \\
\midrule
-- & Rule-Based & 23 / 34 & 67.6 & 0 & 46.4 & 92.9 & 137.3 \\
\midrule
\multirow{2}{*}{GPT-5.2}
& OpenLoop & 28 / 34 & 82.4 & 3.9k & 22.1 & \textbf{44.2} & \textbf{53.6} \\
& Full & 32 / 34 & \textbf{94.1} & 130k & 44.0 & 92.0 & 97.7 \\
\midrule
\multirow{2}{*}{Sonnet 4.6}
& OpenLoop & 26 / 34 & 76.5 & 3.7k & 37.2 & 74.5 & 97.4 \\
& Full & 29 / 34 & \textbf{85.3} & 59k & 14.5 & \textbf{30.1} & \textbf{35.2} \\
\midrule
\multirow{2}{*}{Gemini 3.1}
& OpenLoop & 26 / 34 & 76.5 & 4.3k & 40.3 & 80.6 & 105.4 \\
& Full & 31 / 34 & \textbf{91.2} & 42k & 22.6 & \textbf{45.7} & \textbf{50.1} \\
\bottomrule
\end{tabular}
\end{table*}

Table~\ref{tab:main-results} reports both end-to-end success and estimated monetary cost on VASPBench. In addition to task success, we estimate the total cost of each configuration by combining measured VASP execution time with LLM inference cost. VASP execution is priced at \$2 per H100 GPU-hour, following the cost model in Appendix~\ref{app:execution-cost}, and LLM cost is estimated from total token usage using the official API pricing rates for each backbone, as listed in the table caption.

Across all three backbones, AutoDFT-Full improves task success over AutoDFT-OpenLoop. With GPT-5.2, AutoDFT-Full solves 32 of 34 tasks, achieving the highest overall success rate of 94.1\%, compared with 82.4\% for AutoDFT-OpenLoop and 67.6\% for Rule-Based. This improvement comes with higher execution cost than GPT-5.2 OpenLoop, because the closed-loop system triggers additional recovery, reflection, and rerun steps to rescue difficult tasks. Even in this accuracy-oriented regime, GPT-5.2 Full matches Rule-Based in cost while improving success by 26.5 percentage points.

For Claude Sonnet 4.6 and Gemini 3.1 Pro Preview, closed-loop execution improves both success and cost efficiency. Sonnet Full improves success from 76.5\% to 85.3\% while reducing cost/succ. from \$97.4 to \$35.2. Gemini Full improves success from 76.5\% to 91.2\% while reducing cost/succ. from \$105.4 to \$50.1. These results show that the additional LLM reasoning cost can be more than offset by savings in VASP execution time. In particular, closed-loop monitoring and recovery can terminate unpromising runs earlier, repair recoverable failures, and avoid spending GPU time on workflows whose assumptions have already become invalid.

Overall, Table~\ref{tab:main-results} suggests that closed-loop execution should not be viewed simply as an added LLM overhead. Instead, it trades relatively inexpensive runtime reasoning for more selective use of expensive first-principles computation. Depending on the backbone, this trade-off either improves both success and monetary cost, as in Claude Sonnet 4.6 and Gemini 3.1 Pro Preview, or prioritizes the highest task success at a cost comparable to the rule-based baseline, as in GPT-5.2. More detailed analysis about execution cost can be found in Appendix~\ref{app:execution-cost}.

\textbf{Controlled ablation and statistical robustness.}
We further evaluate OpenLoop, Static-Plan, Static-Plan + LLM Repair, and AutoDFT-Full over three independent runs on all 34 VASPBench tasks using Gemini 3.1 Pro. Their mean success rates and sample standard deviations are 75.5\% $\pm$ 1.7\%, 76.5\% $\pm$ 2.9\%, 81.4\% $\pm$ 4.5\%, and 91.2\% $\pm$ 0.0\%, respectively. AutoDFT-Full consistently solves 31/34 tasks across all three runs, whereas the intermediate baselines exhibit greater run-to-run variation. These results demonstrate that replacing heuristic parameters with LLM-generated upfront parameters provides only marginal improvement, and that adding reactive LLM repair does not close the gap to the full framework. The remaining improvement reflects the combined effects of history-conditioned planning, monitoring, recovery, and reflection, rather than an isolated contribution from any single component.

\textbf{Compute-cost comparison.}
The improved success rate comes with additional LLM inference, but not necessarily a proportional increase in the dominant DFT computational cost. Under Gemini 3.1 Pro, Static-Plan requires approximately 12.4 H100 GPU-hours per task, while Static-Plan + LLM Repair and AutoDFT-Full require 22.2 and 22.6 GPU-hours per task, respectively. The latter two configurations therefore use comparable VASP compute, despite Full achieving a 9.8-percentage-point higher mean success rate. Their respective LLM usages are approximately 9k and 42k tokens per task, corresponding to only about \$0.40 additional inference cost under our pricing model. This comparison highlights the distinction between inexpensive additional reasoning and the substantially larger computational cost of first-principles simulations. Detailed per-run outcomes and resource measurements are provided in Appendix~\ref{app:additional-analysis}.

\subsection{Physical Correctness on Materials Project}
\label{sec:mp-results}

VASPBench evaluates whether a workflow converges and produces validated output files, but task completion alone does not establish that the extracted physical quantities are reliable. We therefore evaluate AutoDFT-Full on a 20-material subset from the Materials Project and compare three commonly used properties, namely the electronic bandgap, total magnetic moment, and formation energy, against Materials Project PBE reference values.

Table~\ref{tab:mp-results} reports property-level MAEs over successfully completed runs. AutoDFT-Full achieves high completion rates across the three backbones, with the number of extracted predictions varying by property. GPT-5.2 completes 16 bandgap, 18 magnetic-moment, and 17 formation-energy predictions; Claude Sonnet 4.6 completes 19 predictions for all properties; and Gemini 3.1 Pro Preview completes 20 bandgap predictions and 19 predictions for both magnetic moment and formation energy. Across completed runs, the MAEs range from 0.26 to 0.80~eV for bandgap, from 0.17 to 0.51~$\mu_B$ for magnetic moment, and from 0.29 to 0.41~eV/atom for formation energy.

These results provide a complementary validation to VASPBench. The benchmark success rates show that AutoDFT can execute diverse DFT workflows to convergence, while the Materials Project comparison shows that the resulting outputs can be parsed into physically meaningful quantities with reasonable agreement to established reference data. This is important because an automation system that merely produces converged calculations without reliable property extraction would be insufficient for downstream materials discovery.

\textbf{Matched-subset comparison.}
Because different backbones successfully complete different subsets of materials, the coverage-aware MAEs in Table~\ref{tab:mp-results} are not directly comparable on identical inputs. We therefore recompute all three properties on the common set of 16 materials for which every backbone produces valid predictions. On this matched subset, GPT-5.2, Claude Sonnet 4.6, and Gemini 3.1 Pro Preview achieve bandgap MAEs of 0.80, 0.30, and 0.46~eV, respectively; magnetic-moment MAEs of 0.17, 0.14, and 0.17~$\mu_B$; and formation-energy MAEs of 0.29, 0.36, and 0.44~eV/atom. The magnetic-moment errors become nearly identical across backbones, whereas the differences in bandgap and formation-energy accuracy persist. This matched comparison controls for differences in evaluation coverage while preserving the original coverage-aware assessment. Further details are provided in Appendix~\ref{app:additional-analysis}.

\begin{table*}[t]
\centering
\caption{Property-level accuracy on the Materials Project subset (20 materials). Values are mean absolute errors against Materials Project PBE reference values, computed only over successfully completed tasks. Lower values are better. $n$ denotes the number of tasks with extracted predictions.}
\label{tab:mp-results}
\begin{tabular}{lccccccc}
\toprule
& \multicolumn{2}{c}{Bandgap (eV)} & \multicolumn{2}{c}{Mag.\ moment ($\mu_B$)} & \multicolumn{2}{c}{Form.\ energy (eV/atom)} \\
\cmidrule(lr){2-3} \cmidrule(lr){4-5} \cmidrule(lr){6-7}
Backbone & $n$ & MAE & $n$ & MAE & $n$ & MAE \\
\midrule
GPT-5.2 & 16 & 0.80 & 18 & \textbf{0.17} & 17 & \textbf{0.29} \\
Claude Sonnet 4.6 & 19 & \textbf{0.26} & 19 & 0.49 & 19 & 0.32 \\
Gemini 3.1 Pro Preview & 20 & 0.41 & 19 & 0.51 & 19 & 0.41 \\
\bottomrule
\end{tabular}
\end{table*}

\subsection{Analysis of Closed-Loop Adaptation}
\label{sec:closed-loop-analysis}

We next analyze the sources of the closed-loop gains observed in Table~\ref{tab:main-results}. We consider two complementary questions: which calculation types benefit most from closed-loop execution, and how often successful AutoDFT-Full runs actually depend on runtime intervention rather than on the initial plan.

\begin{figure*}[ht]
\centering
\includegraphics[width=\textwidth]{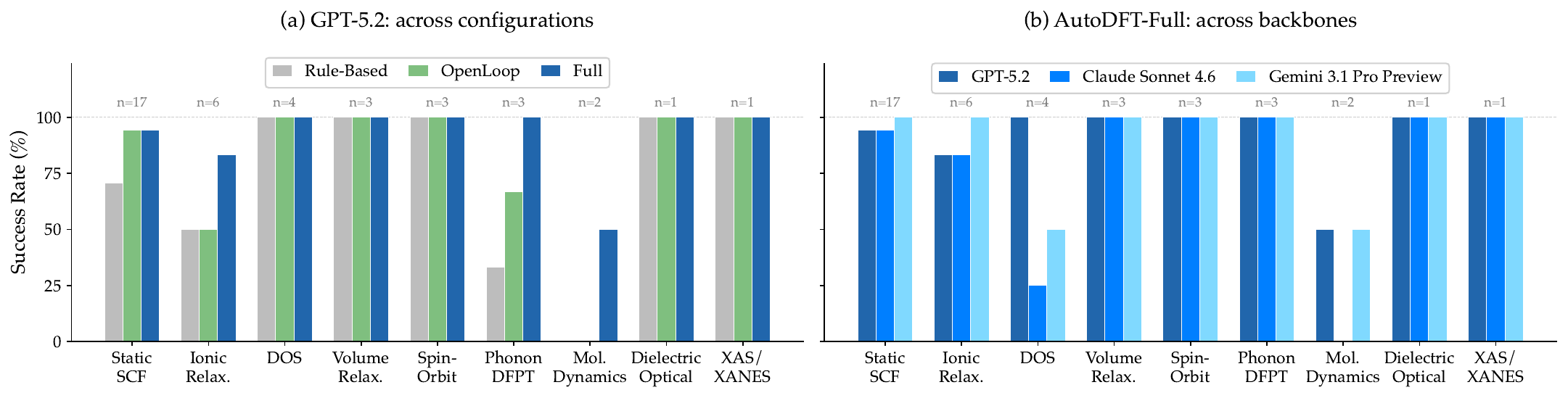}
\caption{Per-calculation-type success rate on VASPBench (34 tasks).
(a)~GPT-5.2 backbone across configurations: the closed-loop components
provide the largest gains on ionic relaxation, phonon DFPT, and molecular dynamics.
(b)~AutoDFT-Full across backbones: performance is consistent on most
types, with the largest variation on DOS and molecular dynamics.
Task counts for each type are shown above the bars.}
\label{fig:per-type}
\end{figure*}

\begin{table*}[ht]
\centering
\caption{Runtime intervention usage among successful AutoDFT-Full runs. Interventions include LLM-driven recovery, monitor-triggered termination, and reflector-driven plan modification. ``w/o intervention'' counts successful runs without these events. Intervention rates are descriptive and do not establish counterfactual task outcomes. In this table, ``Sonnet 4.6'' is short for ``Claude Sonnet 4.6'', and ``Gemini 3.1'' is short for ``Gemini 3.1 Pro Preview''.}
\label{tab:cf-analysis}
\begin{tabular}{llccc}
\toprule
Benchmark & Backbone & Successes & w/o intervention & Intervention rate \\
\midrule
\multirow{3}{*}{VASPBench}
  & GPT-5.2     & 32 & 12 (37.5\%) & 62.5\% \\
  & Sonnet 4.6  & 29 & 22 (75.9\%) & 24.1\% \\
  & Gemini 3.1  & 31 & 29 (93.5\%) &  6.5\% \\
\midrule
\multirow{3}{*}{MP (bandgap)}
  & GPT-5.2     & 16 &  6 (37.5\%) & 62.5\% \\
  & Sonnet 4.6  & 19 & 17 (89.5\%) & 10.5\% \\
  & Gemini 3.1  & 20 & 13 (65.0\%) & 35.0\% \\
\midrule
\multirow{3}{*}{MP (magnetic moment)}
  & GPT-5.2     & 18 &  6 (33.3\%) & 66.7\% \\
  & Sonnet 4.6  & 19 & 16 (84.2\%) & 15.8\% \\
  & Gemini 3.1  & 19 & 15 (78.9\%) & 21.1\% \\
\midrule
\multirow{3}{*}{MP (formation energy)}
  & GPT-5.2     & 17 &  6 (35.3\%) & 64.7\% \\
  & Sonnet 4.6  & 19 & 18 (94.7\%) & 5.3\% \\
  & Gemini 3.1  & 19 & 13 (68.4\%) & 31.6\% \\
\bottomrule
\end{tabular}
\end{table*}

Figure~\ref{fig:per-type} decomposes VASPBench success rates by calculation type. The strongest gains from closed-loop execution occur in multi-stage calculations with nontrivial parameter dependencies. Under GPT-5.2, AutoDFT-Full improves phonon DFPT to 100\% success, compared with 33.3\% for Rule-Based and 66.7\% for OpenLoop, and is the only configuration that solves any molecular dynamics task. It also improves ionic relaxation over fixed templates, where the Rule-Based baseline reaches only 50.0\%. By contrast, simpler or more template-like tasks, such as spin-orbit coupling, volume relaxation, and X-ray absorption spectroscopy, are already solved reliably by Rule-Based. This pattern suggests that closed-loop adaptation is most valuable when a fixed initial plan cannot specify the workflow.

To characterize runtime intervention usage, we inspect the reasoning traces of successful AutoDFT-Full runs and record whether they involve LLM-driven recovery, monitor-triggered termination, or reflector-driven plan modification. Table~\ref{tab:cf-analysis} reports these descriptive intervention statistics across VASPBench and Materials Project (MP)-derived benchmarks. Whereas Figure~\ref{fig:per-type} shows where the largest benchmark-level gains occur, Table~\ref{tab:cf-analysis} measures how often successful runs rely on closed-loop mechanisms during execution. Results show that intervention needs depend on both the backbone model and benchmark. For GPT-5.2, many runs are intervention-dependent across all settings: 62.5\% on VASPBench, 62.5\% on MP bandgap, 66.7\% on MP magnetic moment, and 64.7\% on MP formation energy. These high rates show that runtime interventions frequently occur in successful GPT-5.2 workflows, although the observational analysis alone does not establish whether those interventions were necessary for task completion.

The pattern differs for stronger initial planners. Claude Sonnet 4.6 generally requires fewer interventions, with intervention rates of 24.1\% on VASPBench and 5.6\%–15.8\% on the MP benchmarks. This suggests that a stronger initial plan can reduce the need for runtime correction. However, closed-loop adaptation remains useful even when the initial planner is strong: several successful Sonnet runs still depend on recovery or plan modification. Gemini 3.1 Pro Preview shows a mixed profile. On VASPBench, only 6.5\% of its successful tasks require intervention, suggesting that most successes are achieved without runtime modification. On the MP benchmarks, however, intervention-dependent successes account for 35.0\%, 21.1\%, and 31.6\% of successful runs for bandgap, magnetic moment, and formation energy, respectively. Thus, even with a strong initial planner, closed-loop mechanisms still improve robustness, especially across broader materials-property tasks.

Taken together, the per-type and intervention analyses show that AutoDFT-Full is not merely benefiting from a stronger initial plan. The system gains robustness by detecting failed or questionable intermediate states, revising parameters, terminating unproductive runs, and modifying the workflow when earlier assumptions no longer hold. This mechanism is especially important for phonons, relaxations, and molecular dynamics, where errors often emerge only after partial execution, and it also contributes to MP-derived property prediction tasks where failures can arise from structure-specific convergence, magnetism, or parameter-selection issues. Additional analysis (including execution cost) is reported in Appendix~\ref{app:additional-analysis}, and detailed qualitative examples are provided in Appendix~\ref{app:examples}.

\textbf{Intervention usage versus causal contribution.}
The intervention frequencies in Table~\ref{tab:cf-analysis} describe observed execution behavior and should not be interpreted as proving that individual tasks would have failed without intervention. In particular, history-conditioned Step Planner decisions are part of normal closed-loop execution and are not counted as explicit interventions in this analysis. The intermediate baselines provide complementary evidence by directly comparing alternative execution architectures.

\textbf{Step Reflector ablation.}
To more directly assess the contribution of post-convergence reflection, we additionally evaluate AutoDFT-Full without the Step Reflector using Gemini 3.1 Pro. Removing this component reduces VASPBench success from 31/34 (91.2\%) to 28/34 (82.4\%), corresponding to an 8.8-percentage-point decrease. This result provides direct ablation evidence that evaluating completed calculations and revising subsequent actions can improve end-to-end task success, beyond merely handling explicit VASP execution failures. Together with the Static-Plan comparisons, it supports the benefit of combining result-conditioned planning, recovery, and reflection, while leaving finer-grained attribution among the remaining components for future investigation.

\textbf{Preliminary cross-code portability.}
To examine whether the framework can extend beyond VASP, we implemented a preliminary Quantum ESPRESSO (QE) backend. This adaptation replaces VASP-specific input generation, execution, and output parsing with QE-specific \texttt{pw.x} settings, pseudopotential specifications, \texttt{K\_POINTS}, and output interpretation, while retaining the higher-level planning, history management, and reflection mechanisms. With Gemini 3.1 Pro, all five relatively simple QE tasks completed successfully. Although limited in scale, this experiment provides initial execution-based evidence for the portability of the closed-loop architecture beyond a single DFT package.

\section{Limitations and Ethical Considerations}
\label{sec:limitations}

AutoDFT is a general closed-loop framework, but this study focuses on VASP and evaluates a curated set of VASPBench tasks and a 20-material Materials Project subset. Although a preliminary Quantum ESPRESSO backend has been implemented, comprehensive cross-code validation and broader evaluation across materials, exchange-correlation functionals, and calculation settings remain future work. AutoDFT also inherits the computational cost and applicability limits of DFT, although the additional LLM overhead is small relative to the underlying simulations.

Several limitations warrant further investigation. First, VASPBench contains only 34 tasks, with an uneven distribution across calculation types. Although this design covers diverse VASP functionalities, categories represented by only one or two tasks do not support strong type-specific generalizations. Second, because VASPBench is derived from publicly available VASP documentation, potential exposure to related materials during backbone pretraining cannot be excluded, even though no worked examples or parameter configurations are provided as task inputs. Third, our preliminary Quantum ESPRESSO evaluation covers only five relatively simple tasks, and broader cross-code portability remains to be systematically validated. Finally, the Step Reflector has not yet been evaluated through a large-scale expert-agreement study. An informal inspection of 20 sampled reflector outputs by co-authors with DFT expertise revealed no clear scientific errors, but the judgments were not necessarily equivalent to those of experienced practitioners, and the inspection does not establish expert-level decision quality.

This work involves no human subjects, private data, surveillance, or human-facing decision making. Its main benefit is reducing the expertise barrier and manual effort required for reproducible first-principles workflows. The main risks are erroneous automated outputs and overreliance on them. We mitigate these risks through structured validation, runtime monitoring, auditable execution records, and failure recovery; final scientific conclusions remain subject to domain-expert review.

\section{Generative AI Usage}
\label{sec:genai}

Generative AI tools were used to assist with code development and language polishing during the preparation of this work. All generated or revised content was reviewed and verified by the authors. The authors independently designed the methodology, conducted and validated the experiments, analyzed the results, and determined the scientific claims. The authors take full responsibility for the accuracy, originality, and integrity of the paper and its accompanying code.

\section{Conclusion}

We present AutoDFT, a multi-agent framework that embeds LLM reasoning into every stage of the DFT lifecycle through hierarchical planning decomposition and closed-loop adaptive execution. AutoDFT coordinates seven specialized agents, ranging from purely LLM-driven to purely rule-based, to couple workflow design with result-conditioned parameterization, execution monitoring, failure recovery, and plan revision under emergent findings. On VASPBench (34 tasks from official VASP documentation), AutoDFT achieves 94.1\% success with GPT-5.2, compared to 82.4\% for open-loop planning and 67.6\% for a rule-based baseline, with consistent gains across three model families. Evaluation on a 20-material Materials Project subset confirms that task completion translates into quantitative property accuracy. Comparisons against rule-based and open-loop baselines show that closed-loop adaptation provides gains across model families, with runtime recovery and plan revision proving most valuable for multi-stage calculations involving complex parameter interdependencies.

\bibliographystyle{ACM-Reference-Format}
\bibliography{references}

\clearpage
\appendix
%
%
%

\section{Notation and System Scope}
\subsection{Glossary of DFT, VASP, and Materials-Science Notation}
\label{app:glossary}

To keep the main text concise, we use standard terminology from density-functional theory,
VASP, and computational materials science. This appendix summarizes the acronyms, VASP
file names, and VASP input tags used throughout the paper.

Table~\ref{tab:general_glossary} summarizes the general DFT and materials-science acronyms used throughout the paper. Table~\ref{tab:vasp_file_glossary} summarizes the VASP input files, output files, and intermediate artifacts referenced in the workflow. Table~\ref{tab:vasp_tag_glossary} summarizes the VASP input tags used in workflow planning, runtime monitoring, and error recovery. 

\begin{table*}[t]
\centering
\small
\caption{General DFT and materials-science acronyms used in the paper.}
\label{tab:general_glossary}
\begin{tabularx}{\textwidth}{@{}L{0.12\textwidth} L{0.32\textwidth} >{\raggedright\arraybackslash}X@{}}
\toprule
Acronym & Full name & Meaning in this paper \\
\midrule

SCF & Self-Consistent Field &
Self-consistent electronic calculation; used in static, spin-polarized, and pre-processing steps. \\

NSCF & Non-Self-Consistent Field &
Non-self-consistent calculation, typically performed after an SCF calculation for DOS or band-structure analysis. \\

DOS & Density of States &
Electronic density of states, including total and spin-resolved DOS. \\

PDOS & Projected Density of States &
Density of states projected onto atoms, layers, orbitals, or spin channels. \\

DFPT & Density-Functional Perturbation Theory &
Perturbative DFT method used for phonons, dielectric response, Raman modes, and related properties. \\

SOC & Spin-Orbit Coupling &
Relativistic coupling between spin and orbital degrees of freedom. \\

XAS & X-ray Absorption Spectroscopy &
Spectroscopic calculation type involving X-ray absorption. \\

XANES & X-ray Absorption Near-Edge Structure &
Near-edge region of an X-ray absorption spectrum. \\

PBE & Perdew--Burke--Ernzerhof &
Generalized-gradient exchange-correlation functional commonly used in DFT calculations. \\

GGA & Generalized Gradient Approximation &
Class of exchange-correlation functionals that includes PBE. \\

LDA & Local Density Approximation &
Exchange-correlation approximation based on the local electron density. \\

LSDA+$U$ & Local Spin-Density Approximation plus Hubbard $U$ &
Spin-polarized local-density approximation augmented with a Hubbard-$U$ correction. \\

DFT+$U$ & Density Functional Theory plus Hubbard $U$ &
Hubbard-corrected DFT treatment for localized or strongly correlated electronic states. \\

HF & Hartree--Fock &
Wave-function-based electronic-structure method; also appears in the context of hybrid functionals. \\

HSE & Heyd--Scuseria--Ernzerhof &
Screened hybrid exchange-correlation functional. \\

PBE0 & Perdew--Burke--Ernzerhof 0 &
Hybrid exchange-correlation functional based on PBE. \\

GW & Green's-function $G$ and screened-interaction $W$ approximation &
Many-body perturbation method for quasiparticle electronic structure. \\

NEB & Nudged Elastic Band &
Method for transition-state and reaction-path calculations. \\

MD & Molecular Dynamics &
Time evolution of atomic configurations. \\

AIMD & Ab Initio Molecular Dynamics &
Molecular dynamics driven by first-principles forces. \\

CIF & Crystallographic Information File &
Standard file format for specifying crystal structures. \\

BZ & Brillouin Zone &
Primitive cell in reciprocal space. \\

VBM & Valence Band Maximum &
Highest occupied valence-band energy. \\

CBM & Conduction Band Minimum &
Lowest unoccupied conduction-band energy. \\

FM & Ferromagnetic &
Magnetic ordering with aligned magnetic moments. \\

AFM & Antiferromagnetic &
Magnetic ordering with oppositely aligned magnetic moments. \\

AFM-II & Type-II Antiferromagnetic Ordering &
A common antiferromagnetic ordering pattern, for example in rocksalt NiO. \\

NM & Non-magnetic &
Calculation or reference state without magnetic ordering. \\

TCI & Topological Crystalline Insulator &
Topological insulating phase protected by crystal symmetry. \\

TMD & Transition-Metal Dichalcogenide &
Layered compound class such as MoSe$_2$, PtS$_2$, or HfS$_2$. \\

\bottomrule
\end{tabularx}
\end{table*}

\begin{table*}[t]
\centering
\small
\caption{VASP input files, output files, and intermediate artifacts referenced in the paper.}
\label{tab:vasp_file_glossary}
\begin{tabularx}{\textwidth}{@{}L{0.13\textwidth} L{0.34\textwidth} >{\raggedright\arraybackslash}X@{}}
\toprule
Notation & Full name or description & Meaning in this paper \\
\midrule

INCAR & VASP input file for calculation parameters &
Specifies electronic, ionic, magnetic, convergence, and method-specific settings. \\

KPOINTS & VASP input file for reciprocal-space sampling &
Specifies the $k$-point mesh or path used in reciprocal-space integration. \\

POSCAR & VASP input file for structure &
Specifies lattice vectors, atomic species, and atomic positions. \\

POTCAR & VASP input file for pseudopotentials &
Contains the pseudopotentials used for each atomic species. \\

OUTCAR & VASP main output file &
Contains detailed runtime information, convergence records, errors, forces, stresses, and final properties. \\

OSZICAR & VASP convergence log &
Compact log of electronic and ionic convergence, parsed by the monitor during execution. \\

CONTCAR & VASP relaxed-structure output file &
Stores the final structure after relaxation and is commonly used to initialize downstream steps. \\

CHGCAR & VASP charge-density file &
Stores the charge density; reused in static, DOS, band-structure, or NSCF calculations when appropriate. \\

WAVECAR & VASP wavefunction file &
Stores wavefunction information for restarting or continuing electronic-structure calculations. \\

DOSCAR & VASP density-of-states output file &
Stores total and, when requested, projected density-of-states data. \\

EIGENVAL & VASP eigenvalue output file &
Stores band eigenvalues and occupancies. \\

PROCAR & VASP projected eigenvalue output file &
Stores orbital, atomic, and spin projections of electronic states. \\

XDATCAR & VASP trajectory output file &
Stores atomic trajectories, especially for molecular-dynamics calculations. \\

IBZKPT & VASP irreducible Brillouin-zone $k$-point file &
Stores the symmetry-reduced $k$-point set used internally by VASP. \\

LOCPOT & VASP local-potential output file &
Stores the local electrostatic potential, used for quantities such as work functions. \\

ELFCAR & VASP electron-localization-function output file &
Stores the electron localization function when requested. \\

\bottomrule
\end{tabularx}
\end{table*}

\begin{table*}[t]
\centering
\small
\caption{VASP input tags that appear in the workflow design, monitoring, recovery, or prompt templates.}
\label{tab:vasp_tag_glossary}
\begin{tabularx}{\textwidth}{@{}L{0.13\textwidth} L{0.34\textwidth} >{\raggedright\arraybackslash}X@{}}
\toprule
Tag & Full name or description & Meaning in this paper \\
\midrule

ICHARG & Charge-density initialization tag &
Controls whether VASP initializes, reads, or reuses the charge density, including from CHGCAR. \\

ISTART & Wavefunction restart tag &
Controls whether and how VASP restarts from a WAVECAR file. \\

ISPIN & Spin-polarization tag &
Controls whether the calculation is non-spin-polarized or spin-polarized. \\

MAGMOM & Initial magnetic moments &
Specifies initial magnetic moments for spin-polarized or magnetic calculations. \\

LSORBIT & Spin-orbit-coupling switch &
Enables SOC calculations when set appropriately. \\

ENCUT & Plane-wave energy cutoff &
Controls the kinetic-energy cutoff of the plane-wave basis. \\

PREC & Precision setting &
Controls the numerical precision level used by VASP. \\

ISIF & Ionic relaxation and stress-control tag &
Determines which degrees of freedom are relaxed, such as atomic positions, cell shape, and volume. \\

ISMEAR & Smearing-method tag &
Controls the electronic smearing scheme used for Brillouin-zone integration. \\

SIGMA & Smearing width &
Sets the width of the smearing distribution used with ISMEAR. \\

EDIFF & Electronic convergence threshold &
Sets the stopping criterion for electronic self-consistency. \\

EDIFFG & Ionic convergence threshold &
Sets the stopping criterion for ionic relaxation, typically based on forces or energy changes. \\

NELM & Maximum electronic SCF steps &
Sets the maximum number of electronic iterations allowed in one ionic step. \\

NSW & Maximum ionic steps &
Sets the maximum number of ionic steps for relaxation or molecular dynamics. \\

IBRION & Ionic update or perturbation tag &
Selects the ionic relaxation algorithm, molecular-dynamics mode, or DFPT mode. \\

POTIM & Ionic time-step or relaxation parameter &
Controls the ionic time step in molecular dynamics or the step size in ionic relaxation. \\

ALGO & Electronic minimization algorithm &
Selects the electronic optimization or diagonalization algorithm. \\

LDAU & DFT+$U$ switch &
Enables Hubbard-$U$ corrections for selected atomic species and orbitals. \\

LDAUU & Hubbard-$U$ values &
Specifies the effective Hubbard-$U$ parameters used in DFT+$U$ calculations. \\

LMAXMIX & Charge-mixing angular-momentum cutoff &
Controls the maximum angular-momentum channel included in charge-density mixing. \\

LHFCALC & Hartree--Fock exchange switch &
Enables Hartree--Fock exchange, typically for hybrid-functional calculations. \\

METAGGA & Meta-GGA functional selector &
Specifies a meta-GGA exchange-correlation functional when used. \\

LEPSILON & Dielectric-response switch &
Enables calculation of static dielectric tensors and Born effective charges. \\

LREAL & Real-space projection tag &
Controls whether projection operators are evaluated in real space. \\

ISYM & Symmetry-control tag &
Controls the use of symmetry during the calculation. \\

LCHARG & Charge-density output switch &
Controls whether CHGCAR is written. \\

LWAVE & Wavefunction output switch &
Controls whether WAVECAR is written. \\

\bottomrule
\end{tabularx}
\end{table*}

\subsection{Supported Properties}\label{app:task03-supported-properties}

This appendix summarizes the physical properties currently covered by the system. The supported set includes 30 distinct properties. In addition to the properties listed above, several related targets are explicitly excluded from the current scope: energy above hull, advanced optical excitations beyond the independent-particle response, electronic and ionic transport coefficients, and full phonon dispersions requiring supercell force-constant workflows.

\begin{table*}[t]
\centering
\small
\caption{Supported physical properties and the availability of extraction and evaluation functionality.}
\label{tab:supported-properties}
\begin{tabularx}{\textwidth}{@{}>{\raggedright\arraybackslash}p{0.22\textwidth}>{\raggedright\arraybackslash}X@{}}
\toprule
Property & Description \\
\midrule
Bandgap & Electronic bandgap, including direct or indirect character and band-edge information \\
Band structure & Electronic bands along a high-symmetry reciprocal-space path \\
Density of states & Total and projected electronic density of states \\
Space group & Crystallographic space group inferred from a relaxed structure \\
Magnetic moment & Total magnetic moment for spin-polarized calculations \\
Heat of formation & Formation enthalpy or formation energy per atom \\
Relaxation & Structural relaxation or geometry optimization \\
Total energy & Ground-state DFT total energy \\
Raman spectra & Raman-active vibrational modes and spectra from DFPT calculations \\
Phonons & Zone-center phonon frequencies from DFPT calculations \\
Electronic type & Carrier-type classification, including n-type, p-type, and intrinsic behavior \\
Charged defects & Charged-defect calculations with electron-count adjustment \\
Thermodynamics & Entropy, free energy, heat capacity, and zero-point energy derived from phonons \\
van der Waals correction & Dispersion-corrected DFT calculations, including common semi-empirical and many-body schemes \\
Dielectric response & Static dielectric tensor and Born effective charges \\
Work function & Work function obtained from slab calculations \\
Transition-state search & Nudged elastic band calculations for reaction or diffusion barriers \\
Optical dielectric function & Frequency-dependent dielectric response, refractive index, and related optical quantities \\
Magnetic ordering comparison & Comparison of ferromagnetic, antiferromagnetic, and non-magnetic configurations \\
GW quasiparticles & Quasiparticle electronic structure and bandgap calculations \\
Hybrid functionals & Electronic-structure calculations using hybrid exchange-correlation functionals \\
DFT+U & Hubbard-corrected DFT calculations for correlated materials \\
Spin--orbit coupling & Spin--orbit-coupled calculations, including magnetic anisotropy when applicable \\
Surfaces & Slab and surface calculations, including vacuum regions and dipole corrections \\
Molecules & Isolated atoms or molecules treated in finite simulation cells \\
Adsorption energy & Adsorbate binding or adsorption energies from surface calculations \\
Surface energy & Surface formation or cleavage energies \\
Molecular dynamics & Ab initio molecular dynamics under common thermodynamic ensembles \\
Elastic properties & Elastic constants and derived mechanical moduli at the workflow level \\
Infrared spectra & Infrared-active vibrational modes and spectra from DFPT calculations \\
\bottomrule
\end{tabularx}
\end{table*}

\section{Workflow and Implementation Details}
\label{app:imple}
\subsection{Workflow Controller Pseudocode}
\label{app:algorithm}

The pseudocode of AutoDFT workflow controller is shown in Algorithm~\ref{alg:workflow}.

\begin{algorithm}[ht]
\caption{AutoDFT Workflow Controller}
\label{alg:workflow}
\small
\begin{algorithmic}[1]
\Require Crystal structure $\mathcal{S}$, task description $\mathcal{T}$, token budget $B$
\Ensure Computed properties $\mathcal{P}$, reasoning log $\mathcal{L}$

\State $\text{plan} \leftarrow \text{StrategicPlanner}(\mathcal{S}, \mathcal{T})$ \Comment{Skeletal plan}
\State $\text{ctx} \leftarrow \text{ExecutionContext}(\text{plan})$
\State $n_{\text{rev}} \leftarrow 0$ \Comment{Plan revision counter}

\While{$\text{ctx.step\_index} < |\text{plan.steps}|$}
  \State $s \leftarrow \text{plan.steps[ctx.step\_index]}$
  \State $\theta \leftarrow \text{StepPlanner}(s, \text{ctx})$ \Comment{INCAR/KPOINTS}
  \State $\text{job} \leftarrow \text{VASPExecutor.submit}(\theta)$

  \For{$k_{\text{retry}} = 1, \ldots, K_{\text{retry}}$}
    \State $r \leftarrow \text{Monitor.watch}(\text{job}, \text{ctx})$ \Comment{Dual-path monitoring}
    \If{$r.\text{status} = \textsc{Converged}$}
      \State \textbf{break}
    \ElsIf{$r.\text{status} = \textsc{Terminate}$}
      \State $\theta \leftarrow \text{Recovery}(\theta, r, \text{ctx})$ \Comment{Modified parameters}
      \State $\text{job} \leftarrow \text{VASPExecutor.submit}(\theta)$
    \EndIf
  \EndFor

  \State $\text{outcome} \leftarrow \text{ParseOutputs}(\text{job})$
  \State $\text{ctx.update}(\text{outcome})$
  \State $d \leftarrow \text{StepReflector}(\text{outcome}, \text{plan}, \text{ctx})$

  \If{$d.\text{decision} = \textsc{Accept}$}
    \State $\text{ctx.step\_index} \mathrel{+}= 1$
  \ElsIf{$d.\text{decision} = \textsc{Redo}$}
    \State \textbf{continue} \Comment{Retry same step}
  \ElsIf{$d.\text{decision} = \textsc{ModifyPlan}$ \textbf{and} $n_{\text{rev}} < R_{\max}$}
    \State $\text{plan} \leftarrow d.\text{modified\_plan}$
    \State $n_{\text{rev}} \mathrel{+}= 1$
  \Else
    \State $\text{ctx.step\_index} \mathrel{+}= 1$ \Comment{Accept on budget exhaustion}
  \EndIf
\EndWhile

\State $\mathcal{P} \leftarrow \text{PostprocessingAgent}(\text{ctx})$
\Return $\mathcal{P}, \text{ctx.reasoning\_log}$
\end{algorithmic}
\end{algorithm}

\subsection{Basic Implementation Details}
\label{app:implementation-details}

All calculations are run using VASP 6.4.0 compiled with the NVIDIA HPC SDK and FFTW. Each run uses a single NVIDIA H100 GPU. The token budget for each workflow is capped at 200{,}000 tokens, and runs that exhaust this budget are recorded as failures. The maximum number of recovery attempts per failed step is set to 3. The maximum number of workflow steps is capped at 15 to prevent plan inflation. Each individual VASP step has a wall-clock timeout of 4 hours. Steps exceeding this limit are treated as failures rather than skipped. Each configuration is run once per task.

All VASP calculations in this work were performed under a valid institutional VASP license. We do not redistribute VASP binaries, POTCAR files, pseudopotentials, or any other license-restricted VASP assets. The released benchmark contains task specifications, structures, and prompts sufficient to reproduce the benchmark setup, but users are responsible for obtaining and using VASP and any required pseudopotentials under their own licenses.
\subsection{Rule-Based Monitor Detection Rules}
\label{app:task04-rule-monitor}

This section summarizes the detection rules used by the rule-based monitoring component. The monitor combines OSZICAR-based heuristics, including electronic divergence, oscillatory relaxation, stagnation, and abnormal physical indicators, with a regular-expression bank applied to OUTCAR to identify explicit VASP error messages. The rules are adapted for density-functional perturbation theory calculations, spin-orbit-coupled calculations, and static calculations, where the interpretation of ionic and electronic progress differs from standard structural relaxations.

\begin{table*}[t]
\centering
\small
\caption{Default monitor configuration.}
\label{tab:rule-monitor-config}
\begin{tabularx}{\textwidth}{@{}>{\raggedright\arraybackslash}p{0.31\textwidth} >{\raggedright\arraybackslash}X@{}}
\toprule
Parameter & Default \\
\midrule
\texttt{nelm\_threshold} & 60 \\
\texttt{oscillation\_window} & 10 ionic steps \\
\texttt{stagnation\_de\_threshold} & $1\times 10^{-4}$ eV \\
\texttt{stagnation\_step\_threshold} & 20 ionic steps \\
\texttt{force\_threshold} & 100.0 eV/\AA \\
\texttt{energy\_threshold} & 0.0 eV, used as a positive-energy warning threshold \\
\texttt{check\_interval} & 600 s, corresponding to 10 min \\
\texttt{soc\_milestone\_interval} & Every 10 electronic steps, used to reset the timeout counter \\
\texttt{positive\_energy\_grace\_steps} & 3 ionic steps \\
\bottomrule
\end{tabularx}
\end{table*}

\begin{table*}[t]
\centering
\small
\caption{OSZICAR-based heuristic checks.}
\label{tab:rule-monitor-heuristics}
\begin{tabularx}{\textwidth}{@{}
>{\raggedright\arraybackslash}p{0.27\textwidth}
>{\raggedright\arraybackslash}X@{}}
\toprule
Detection type & Trigger condition \\
\midrule
\texttt{ELECTRONIC\_DIVERGENCE}
& The last three consecutive ionic steps each require at least \texttt{NELM} electronic steps, with the default threshold set to 60. \\

\texttt{ENERGY\_OSCILLATION}
& Within the most recent 10 ionic steps, the number of sign changes in $dE$ is at least 5, and either the net energy change is non-negative or $|\Delta E|$ is smaller than the average $|dE|$. \\

\texttt{FORCE\_STAGNATION}
& At least 16 of the most recent 20 ionic steps have $|dE| < 10^{-4}$ eV, while the maximum force remains larger than $|\texttt{EDIFFG}|$. \\

\texttt{POSITIVE\_ENERGY}
& The most recent ionic-step energy is positive after the three-step grace period. \\

\texttt{EXTREME\_FORCES}
& The maximum force exceeds 100 eV/\AA. \\
\bottomrule
\end{tabularx}
\end{table*}

\begin{table*}[t]
\centering
\small
\caption{OUTCAR-based error checks.}
\label{tab:rule-monitor-regex}
\begin{tabularx}{\textwidth}{
>{\raggedright\arraybackslash}p{0.27\textwidth}
>{\raggedright\arraybackslash}X}
\toprule
Detection type & OUTCAR diagnostic criterion \\
\midrule
\texttt{WAVECAR\_BASIS\_CHANGE}
& Detects restart incompatibilities caused by changes in the plane-wave basis associated with the stored WAVECAR. \\

\texttt{ZBRENT\_ERROR}
& Detects failures of the ZBRENT line-search procedure during ionic relaxation. \\

\texttt{EDDDAV\_ERROR}
& Detects electronic diagonalization failures associated with the EDDDAV solver. \\

\texttt{BRMIX\_ERROR}
& Detects severe charge-mixing instabilities during self-consistent-field iterations. \\

\texttt{GRID\_ERROR}
& Detects failures associated with FFT grids, real-space grids, or grid-consistency checks. \\

\texttt{PRICEL\_ERROR}
& Detects symmetry-cell construction failures or non-integer symmetry-related elements. \\

\texttt{CHGCAR\_READ\_ERROR}
& Detects charge-density restart failures caused by unreadable or incompatible CHGCAR data. \\

\texttt{MAGMOM\_ERROR}
& Detects failures in parsing or validating the magnetic-moment specification. \\

\texttt{HYBRID\_ICHARG\_CONFLICT}
& Detects incompatible restart settings for hybrid-functional calculations involving Fock exchange and fixed charge-density modes. \\

\texttt{ADDGRID\_TS\_CONFLICT}
& Detects unsupported combinations of additional grid settings and Tkatchenko--Scheffler dispersion corrections. \\

\texttt{KPOINTS\_MODE\_CONFLICT}
& Detects incompatible k-point sampling modes, especially conflicts between line-mode paths and calculations requiring uniform meshes. \\

\texttt{NOMEGA\_LIMIT}
& Detects cases where the requested number of frequency-grid points exceeds the supported limit. \\

\texttt{SYMPREC\_ERROR}
& Detects symmetry-tolerance, lattice-consistency, or symmetry-mapping failures. \\

\texttt{INSUFFICIENT\_BANDS}
& Detects cases where the highest electronic band remains occupied, indicating an insufficient number of bands. \\

\texttt{MEMORY\_ERROR}
& Detects memory-allocation failures, out-of-memory conditions, or process termination caused by insufficient memory. \\

\texttt{EDDDAV\_SUBSPACE\_ERROR}
& Detects non-Hermitian subspace-matrix failures during electronic minimization. \\
\bottomrule
\end{tabularx}
\end{table*}

\paragraph{Special-case calculation handling.}
For density-functional perturbation theory calculations with \texttt{IBRION=7} or \texttt{IBRION=8}, standard ionic-step monitoring is disabled because the calculation proceeds through symmetry-reduced perturbations rather than ordinary ionic updates. Progress markers of the form ``Degree of freedom: $N/M$'' are instead used to identify new perturbations, and each new perturbation resets the timeout counter. This treatment is used for phonon, Raman, and infrared calculations.

For spin-orbit-coupled calculations with \texttt{LSORBIT=.TRUE.}, static calculations may still be computationally expensive because of the spinor basis and the corresponding increase in band count. The monitor tracks the spin-axis direction, resets the timeout counter upon SCF convergence, and also resets it after every \texttt{soc\_milestone\_interval} electronic steps. When OSZICAR output is empty or delayed, the electronic progress is supplemented using standard output streams.

For static calculations with \texttt{NSW=0} or \texttt{IBRION=-1}, only electronic SCF progress is monitored. The timeout counter is reset every 20 electronic steps to avoid prematurely terminating long but still progressing calculations.

Convergence is detected from the presence of ``reached required accuracy'' in the trailing portion of OUTCAR. For static calculations, the completion marker ``General timing and accounting informations'' is also treated as evidence that the calculation has finished normally. These rules correspond to the rule-based monitoring procedure described in the main text and to the failure-detection rules summarized in this appendix.

\subsection{Recovery Procedure}\label{app:task05-recovery}

This section summarizes the recovery protocol used after a failed VASP calculation. The recovery mechanism is designed to preserve diagnostic information from the failed run, construct a structured failure context, and generate a revised calculation setup for the subsequent attempt.

\begin{table*}[t]
\centering
\small
\caption{Default recovery configuration.}
\label{tab:recovery-config}
\begin{tabularx}{\textwidth}{@{}>{\raggedright\arraybackslash}p{0.31\textwidth}>{\raggedright\arraybackslash}X@{}}
\toprule
Configuration field & Default setting \\
\midrule
archive naming pattern & \texttt{attempt\_\{attempt\}\_\{reason\}} with timestamp appended \\
maximum retry attempts & 3 \\
preserve WAVECAR & False \\
preserve CHGCAR & False \\
restart from CONTCAR & True \\
archive all outputs & True \\
\bottomrule
\end{tabularx}
\end{table*}

For each failed calculation, diagnostic and output files are moved to a dedicated archive directory inside the calculation directory. The directory name records the retry index, the detected failure category, and a timestamp. By default, the recovery procedure archives the main VASP output files, including \texttt{OUTCAR}, \texttt{OSZICAR}, \texttt{CONTCAR}, \texttt{CHGCAR}, \texttt{WAVECAR}, \texttt{vasprun.xml}, \texttt{DOSCAR}, \texttt{EIGENVAL}, \texttt{PROCAR}, \texttt{PCDAT}, \texttt{XDATCAR}, \texttt{IBZKPT}, \texttt{CHG}, \texttt{LOCPOT}, and \texttt{ELFCAR}, together with scheduler logs and run reports when present. When full-output archiving is enabled, the corresponding input files, \texttt{INCAR}, \texttt{POSCAR}, \texttt{KPOINTS}, and \texttt{POTCAR}, are archived as well. After archiving, the next attempt is initialized using the archived input state, with \texttt{POSCAR} replaced by \texttt{CONTCAR} when restart from the final ionic configuration is enabled.

The recovery model receives a structured representation of the failed calculation, including the detected failure category, the suggested recovery intent, the recent electronic-convergence trajectory from \texttt{OSZICAR}, a compact summary of \texttt{OUTCAR}, and the current \texttt{INCAR} and \texttt{KPOINTS} settings. This context is augmented with a domain knowledge excerpt covering common VASP failure modes and parameter-level remedies. The model then proposes a revised calculation setup through a constrained structured output format.

The required recovery output contains either a complete replacement of the \texttt{INCAR} settings or a patch-style set of \texttt{INCAR} overrides, and analogously either a complete replacement of the \texttt{KPOINTS} settings or a patch-style set of \texttt{KPOINTS} overrides. The output also specifies the requested VASP executable variant, a confidence score, a concise rationale for the proposed changes, and optional information about whether the original calculation plan should be modified. The structured interface is intended to make recovery decisions auditable while preventing unconstrained free-form edits to the calculation inputs.

Formally, the recovery output schema includes the following fields:
\begin{itemize}
\item \texttt{updated\_parameters.incar} or \texttt{incar\_overrides};
\item \texttt{updated\_parameters.kpoints} or \texttt{kpoints\_overrides};
\item \texttt{vasp\_version}, selected from the supported executable variants;
\item \texttt{confidence}, a scalar score between 0 and 1;
\item \texttt{reasoning\_trace}, a concise explanation of the recovery decision;
\item \texttt{should\_modify\_plan} and \texttt{plan\_revision\_note}, used when the failed calculation indicates that the broader workflow plan should be revised.
\end{itemize}

The revised inputs are validated before the next calculation attempt. Invalid or incomplete recovery outputs are rejected, ensuring that only well-formed parameter updates are applied. The prompt template and output schema used by the recovery model are provided in Appendix~\ref{app:task07-prompts}.

\subsection{LLM Prompt Templates}\label{app:task07-prompts}

This section reproduces the system and user prompt templates that drive the AutoDFT agents.

\renewcommand\lstlistingname{Prompt}

\subsubsection{VASP Knowledge Base Constant Used by the Step Planner and Recovery Prompts}
\leavevmode\par

\begin{lstlisting}[caption={VASP\_KNOWLEDGE\_BASE constant.}]
VASP parameter guidance:
- Use `ISPIN=2' and explicit `MAGMOM' for materials likely containing
  open-shell transition metals.
- For 2D materials prefer `ISIF=2'; for bulk relaxations prefer `ISIF=3'.
- Metallic systems generally use `ISMEAR=1 or 2'; semiconductors use
  `ISMEAR=0' and small `SIGMA'.
- Difficult convergence can require `ALGO=Normal' or `ALGO=All',
  smaller `POTIM', and conservative mixing.
- SOC or non-collinear calculations require `vasp_ncl'.
- Dense k-mesh and tighter `EDIFF' belong in static/band/DOS steps,
  not cheap pre-relaxation.
\end{lstlisting}

\subsubsection{Strategic Planner}\label{app:prompts-strategic}
\leavevmode\par

\begin{lstlisting}[caption={Strategic Planner system prompt.}]

You are a DFT workflow strategist. Output JSON only. Produce a skeletal plan: decide step sequence, objectives, constraints, success criteria, and scientific rationale. Do not choose detailed INCAR or KPOINTS values unless they are unavoidable hard constraints.

CRITICAL PLANNING CONSTRAINTS:
1. EFFICIENCY: Each step must advance the calculation toward the goal. Do NOT
   include pure validation, verification, integrity-check, or checkpoint
   steps that run a VASP calculation solely to confirm outputs are
   parseable. Sanity checks belong as success_criteria within computation
   steps, not as separate steps. Target 3-6 steps for standard tasks
   (relaxation, bandgap, DOS, elastic constants). Only exceed this for
   genuinely multi-stage methods (GW, NEB, DFPT).
2. SINGLE-STRUCTURE EXECUTION: The execution engine runs exactly ONE VASP
   calculation per step. If a workflow requires calculations at multiple
   structures (e.g., EOS fitting with N volume points), each volume point
   must be its own step. When only the equilibrium volume is needed, prefer
   ISIF=3 variable-cell relaxation over manual EOS fitting -- it is cheaper
   and sufficient for most purposes.
3. NO REDUNDANT SINGLE-POINTS: Do not add a preliminary single-point
   calculation before a relaxation just to "establish a baseline" -- the
   first ionic step of the relaxation already provides this information.

COMPUTE BUDGET AWARENESS:
- Each VASP step has a hard wall-time limit of 4 hours. If a step cannot
  realistically finish one ionic iteration (or one SCF) within that budget,
  the whole workflow fails and no result is reported. Plan accordingly.
- Relative costs (rough rule of thumb, ceteris paribus):
    PBE / PBEsol / LDA           1x   (baseline)
    DFT+U                       ~1x
    SOC (LSORBIT=.TRUE.)        ~3-4x (spinor basis; roughly doubles memory)
    meta-GGA (SCAN, TPSS, MBJ)  ~2-3x
    Hybrid (HSE, PBE0, B3LYP)   ~30-100x
    Hybrid + SOC                ~100-300x
    GW                          ~1000x and up
- Concrete implications:
    * For bandgap tasks: the default is PBE (+ SOC only when heavy elements
      demand it, e.g. Bi/Pb/Hg/Tl/I/Te in systems where SOC qualitatively
      changes the gap). A three-step workflow (relax -> SCF -> NSCF band
      path) is usually the right answer.
    * Do NOT tack on "optional" HSE / hybrid / mBJ refinement steps "for
      higher accuracy" unless the user task explicitly demands it.
    * If the user task truly requires hybrid-level accuracy, keep the
      workflow short (relax-PBE -> HSE-SCF on a coarse k-mesh), rather
      than stacking PBE + SOC + HSE + HSE-SOC on top of each other.
4. CONDITIONAL STEPS ARE USUALLY EXECUTED: the orchestrator does not have
   a robust "skip if not needed" check for free-form `condition' strings,
   so any step included in the plan should be expected to actually run.
   Prefer leaving optional refinement steps out of the plan entirely and
   letting the reflector add them on evidence of need, rather than listing
   them up front as `condition: "run only if ..."'.
\end{lstlisting}

\begin{lstlisting}[caption={Strategic Planner user prompt template.}]
Task: {task_description}
Material ID: {material_id}
Formula: {formula}
Material type: {material_type}
Additional context: {additional_context or 'None'}

POSCAR:
{poscar_content}

IMPORTANT: Aim for at most 6 steps. Every step must run a meaningful VASP
calculation that advances toward the goal. Do not add separate verification,
integrity-check, or baseline-confirmation steps.

Return JSON with keys:
- task_description
- material_formula
- material_type
- overall_strategy
- global_considerations
- reasoning_trace
- steps: list of objects with step_name, objective, computation_type,
  depends_on, constraints, success_criteria, priority, condition
\end{lstlisting}

\subsubsection{Step Planner}\label{app:prompts-step}
\leavevmode\par

\begin{lstlisting}[caption={Step Planner system prompt.}]
You are a VASP step planner.
Output JSON only with concrete VASP parameters for one step.
Ground decisions in prior step outcomes and the step objective.
If uncertainty is high, include fallback parameters and say why.
\end{lstlisting}

\begin{lstlisting}[caption={Step Planner user prompt template.}]
Knowledge base:
{VASP_KNOWLEDGE_BASE}

Material context:
{material_context as JSON}

Current skeletal step:
{current_skeletal_step as JSON}

Prior outcomes:
{formatted prior step outcomes}

Prior step inputs (actual parameters that were written to disk):
{formatted prior step inputs
 -- shows only cross-step-sensitive keys: ENCUT, PREC, ISPIN, LSORBIT,
    LNONCOLLINEAR, LHFCALC, METAGGA, GGA, LDAU, LMAXMIX}

CROSS-STEP CONSISTENCY NOTES:
- When this step sets ICHARG >= 10 (e.g. 11 for NSCF band structures, 1
  for restart), it reads the CHGCAR produced by an earlier step. VASP
  requires the FFT grid to match between the CHGCAR source and the current
  step, which means ENCUT and PREC must be identical to the source step.
  If you omit ENCUT or PREC, VASP derives them from POTCAR defaults and
  the grid may differ, causing "charge density could not be read from file
  CHGCAR for ICHARG>10". When in doubt, copy ENCUT and PREC verbatim from
  the SCF step that wrote the CHGCAR.
- ISPIN, LSORBIT, LNONCOLLINEAR, LHFCALC, METAGGA, and LDAU/LDAUU must
  also stay consistent with the CHGCAR / WAVECAR source unless you are
  deliberately changing the electronic framework -- doing so generally
  requires a fresh SCF (ICHARG=2) rather than an NSCF restart.
- For non-collinear (LSORBIT=.TRUE.) steps, a collinear WAVECAR cannot be
  reused (spinor basis doubles the coefficients). Start from the CHGCAR
  only (ISTART=0, ICHARG=1) unless a prior SOC step produced a compatible
  WAVECAR.

Return JSON with keys:
- incar
- kpoints
- additional_files
- vasp_version
- reasoning_trace
- confidence
- fallback_parameters
\end{lstlisting}

\subsubsection{Reasoning Monitor}\label{app:prompts-monitor}
\leavevmode\par

\begin{lstlisting}[caption={Reasoning Monitor system prompt.}]
You are monitoring a running VASP calculation.
Output JSON only. Decide one of: continue, converged, intervene, terminate.
Prefer conservative actions and cite the observed convergence behavior.
\end{lstlisting}

\begin{lstlisting}[caption={Reasoning Monitor user prompt template.}]
Analyze this running calculation state:
{monitor_context as JSON
 -- contains: OSZICAR tail, electronic-history list,
    elapsed time, current INCAR snapshot}

Return JSON with keys:
- decision
- confidence
- reasoning_trace
- should_kill
- suggested_action
\end{lstlisting}

\subsubsection{Reasoning Recovery}\label{app:prompts-recovery}
\leavevmode\par

\begin{lstlisting}[caption={Reasoning Recovery system prompt.}]
You are recovering a failed VASP calculation.
Output JSON only. Diagnose the likely cause, propose parameter/file changes,
and keep the fix minimal but technically justified.

When POSITIVE_ENERGY is reported, reason step by step:
1. What is the electronic structure of this material? Metal, semiconductor,
   or insulator?
2. Check ISMEAR in the current INCAR. ISMEAR=1 or 2 (Methfessel-Paxton /
   Fermi) is designed for metals. Using it on a semiconductor or insulator
   can produce unphysically large entropy contributions and positive total
   energies. If the material is NOT a metal, ISMEAR should be 0 (Gaussian
   smearing) with SIGMA <= 0.05 eV.
3. Check for atomic overlaps: are any interatomic distances < 1.5 Angstrom?
   If so, the geometry itself is unphysical and must be fixed (scale or
   perturb POSCAR) before changing INCAR.
4. Only after ruling out geometry issues and smearing mismatch should you
   adjust convergence parameters (ALGO, AMIX, NELM, etc.).
\end{lstlisting}

\begin{lstlisting}[caption={Reasoning Recovery user prompt template.}]
Knowledge base:
{VASP_KNOWLEDGE_BASE}

Recovery context:
{recovery_context as JSON
 -- contains: failure_reason, suggested_fix, last 30 OSZICAR lines,
    OUTCAR error excerpt, current INCAR/KPOINTS, attempt number}

Return JSON with keys:
- updated_parameters: dict with the following fields:
    - "incar" OR "incar_overrides" (mutually exclusive, exactly one must
      be provided):
        - "incar": full replacement INCAR dict (non-null, non-empty)
        - "incar_overrides": partial patch dict with only the keys to change
      Providing both, or neither (both null/missing), is an error.
    - "kpoints": full kpoints string/dict, or null to keep unchanged
    - "kpoints_overrides": partial kpoints patch, or null;
      mutually exclusive with "kpoints"
    - "vasp_version": string or null to keep unchanged
    - "confidence": float 0-1 or null
    - "additional_files": dict or null
- reasoning_trace: string explaining the diagnosis and fix
- should_modify_plan: bool
- plan_revision_note: string or null
\end{lstlisting}

\subsubsection{Step Reflector}\label{app:prompts-reflector}
\leavevmode\par

\begin{lstlisting}[caption={Step Reflector system prompt.}]
You are reflecting on a completed DFT step.
Output JSON only. Decide one of: accept, redo, modify_plan.
Judge physical plausibility and whether the downstream plan should change.
\end{lstlisting}

\begin{lstlisting}[caption={Step Reflector user prompt template.}]
Reflection context:
{reflection_context as JSON
 -- contains: just-completed StepOutcomeSummary,
    current SkeletalPlan, prior outcomes}

Return JSON with keys:
- decision: "accept" | "redo" | "modify_plan"
- reasoning_trace: string explaining your reasoning
- modified_plan: null if decision != "modify_plan"; otherwise a COMPLETE
  plan dict with ALL of the following keys (copy unchanged fields verbatim
  from current_skeletal_plan):
    {
      "task_description": string,
      "material_formula": string,
      "material_type": string,
      "overall_strategy": string,
      "steps": [...],
      "global_considerations": [...],
      "reasoning_trace": string,
      "material_id": string,
      "metadata": {}
    }
- next_step_index: int (index of next step to execute in the new or
  current plan)
\end{lstlisting}

\section{Benchmark Details}
\subsection{VASPBench Benchmark Task List}\label{app:task01-vaspbench}

\begin{table}[t]
\centering
\small
\caption{Distribution of calculation types within the 34-task subset. Materials with multiple ground-truth configurations are counted more than once.}
\label{tab:vaspbench-calctype-counts}
\begin{tabular}{@{}lr@{}}
\toprule
Calculation type & Count \\
\midrule
Static SCF & 17 \\
Ionic relaxation & 6 \\
DOS & 4 \\
Volume relaxation & 3 \\
Spin orbit & 3 \\
Phonon DFPT & 3 \\
Molecular dynamics & 2 \\
Dielectric optical & 1 \\
XAS XANES & 1 \\
\bottomrule
\end{tabular}
\end{table}

This section enumerates all 34 systems in the VASPBench evaluation subset, along with their stoichiometric composition and the calculation types as recorded in the ground-truth database.

The distribution is consistent with the statistics cited in the main text ($n=17$ for static SCF entries, and groups of 2--6 for relaxation, SOC, DFPT, and DOS tasks). The total number of entries is 34.

Space-group information is sparsely recorded in the ground-truth database because most entries correspond to molecular, surface, or non-bulk systems. Only three entries have a recorded space group, namely space group 225, Fm$\bar{3}$m. Target properties are determined by the corresponding ground-truth calculation type. For example, DOS-type entries are associated with density-of-states targets, whereas static SCF entries provide total energies and band gaps derived from eigenvalue spectra.

\begin{table*}[t]
\centering
\small
\caption{Per-task summary of VASPBench.}\label{tab:vaspbench-pertask}
\begin{tabularx}{\textwidth}{@{}r L{0.38\textwidth} L{0.13\textwidth} >{\raggedright\arraybackslash}X@{}}
\toprule
\# & Example ID & Composition & Calculation\ Type \\
\midrule
1 & Adsorption of H$_2$O on TiO$_2$ & Ti$_2$O$_5$H$_2$ & Ionic relaxation, molecular dynamics \\
2 & Alpha-AlF$_3$ & Al$_2$F$_6$ & DOS \\
3 & Alpha-SiO$_2$ & Si$_3$O$_6$ & DOS \\
4 & Bandgap of Si using different DFT+HF methods & Si$_2$ & Static SCF \\
5 & Beta-tin Si & Si$_2$ & Volume relaxation \\
6 & Calculate U for LSDA+U & Ni$_{16}$O$_{16}$ & Static SCF \\
7 & Cd Si relaxation & Si$_2$ & Ionic relaxation \\
8 & Cd Si volume relaxation & Si$_2$ & Volume relaxation \\
9 & CO & CO & Ionic relaxation \\
10 & CO partial DOS & CO & Static SCF \\
11 & CO vibration & CO & Phonon DFPT \\
12 & Constraining local magnetic moments & Fe$_2$ & Static SCF \\
13 & Constraining the local magnetic moments & Ni$_2$O$_2$ & Spin orbit \\
14 & Estimation of J magnetic coupling & Ni$_2$O$_2$ & Static SCF \\
15 & Fcc Ni & Ni & Static SCF, DOS \\
16 & Fcc Ni (revisited) & Ni & Static SCF \\
17 & Fcc Ni DOS & Ni & Static SCF \\
18 & Fcc Si & Si & Static SCF \\
19 & Fcc Si DOS & Si & Static SCF \\
20 & Graphite interlayer distance & C$_4$ & Volume relaxation \\
21 & H$_2$O & H$_2$O & Ionic relaxation \\
22 & H$_2$O vibration & H$_2$O & Phonon DFPT \\
23 & Including the Spin-Orbit Coupling & Ni$_2$O$_2$ & Spin orbit \\
24 & Ionic contributions: NaCl & NaCl & Dielectric optical, phonon DFPT, static SCF \\
25 & Ni(100) surface DOS & Ni$_5$ & Static SCF, DOS \\
26 & Ni(111) surface relaxation & Ni & Ionic relaxation, static SCF \\
27 & NiO LSDA+U & Ni$_2$O$_2$ & Static SCF \\
28 & Nucleophile Substitution CH$_3$Cl - Standard MD & CH$_3$Cl$_2$ & Molecular dynamics \\
29 & O atom & O & Static SCF \\
30 & O atom spinpolarized & O & Static SCF \\
31 & O atom spinpolarized low symmetry & O & Static SCF \\
32 & O dimer & O$_2$ & Ionic relaxation \\
33 & Spin-orbit coupling in a Ni monolayer & Ni & Spin orbit \\
34 & XANES in Diamond & C$_{128}$ & XAS XANES \\
\bottomrule
\end{tabularx}
\end{table*}

\subsection{Materials Project 20-Material Benchmark}\label{app:task11-mp20}

This section documents the curated 20-material Materials Project subset used as the multi-property benchmark. Selection criteria include at most 24 sites per unit cell, four experimental band-gap categories, chemical and physical diversity, and the availability of PBE reference values for band gap, formation energy per atom, and magnetic moment. The final benchmark contains five metals, five small-gap materials, five medium-gap materials, and five large-gap materials. It spans oxides, chalcogenides, halides, pnictides, intermetallics, Heusler alloys, perovskites, transition-metal dichalcogenides, and fluorides. Three materials have magnetic ground states, namely ferromagnetic MnNi, ferromagnetic GdSb, and antiferromagnetic NiO, while the remaining 17 materials are non-magnetic.

\begin{table*}[t]
\centering
\small
\caption{20-material Materials Project benchmark.}
\label{tab:mp20-materials}
\begin{tabular}{llllc}
\toprule
mp-id & Formula & Electronic class & Magnetic ordering & Magnetic? \\
\midrule
mp-21075 & HfC & metal & NM & $\times$ \\
mp-11500 & MnNi & metal & FM (4.36 $\mu_B$) & $\checkmark$ \\
mp-510403 & GdSb & metal & FM (7.03 $\mu_B$) & $\checkmark$ \\
mp-1885 & AlIr & metal & NM & $\times$ \\
mp-10910 & Al$_2$Ru & metal & NM ($\sim 0$) & $\times$ \\
mp-2730 & HgTe & small gap (SOC) & NM & $\times$ \\
mp-1883 & SnTe & small gap (TCI) & NM & $\times$ \\
mp-5967 & TiCoSb & small gap (half-Heusler) & NM ($\sim 0$) & $\times$ \\
mp-762 & PtS$_2$ & small gap (TMD) & NM & $\times$ \\
mp-1634 & MoSe$_2$ & small gap (TMD) & NM & $\times$ \\
mp-2490 & GaP & medium gap (III--V) & NM & $\times$ \\
mp-23231 & AgBr & medium gap (DFT undershoot) & NM & $\times$ \\
mp-361 & Cu$_2$O & medium gap (cuprite) & NM ($\sim 0.012$) & $\times$ \\
mp-985829 & HfS$_2$ & medium gap (TMD) & NM & $\times$ \\
mp-1779 & YbTe & medium gap (4f) & NM ($\sim 0$) & $\times$ \\
mp-19009 & NiO & large gap (Mott) & AFM & $\checkmark$ \\
mp-856 & SnO$_2$ & large gap (rutile) & NM ($\sim 0$) & $\times$ \\
mp-3614 & KTaO$_3$ & large gap (perovskite) & NM & $\times$ \\
mp-2741 & CaF$_2$ & large gap (fluorite) & NM & $\times$ \\
mp-22939 & BiClO & large gap (oxyhalide) & NM & $\times$ \\
\bottomrule
\end{tabular}
\end{table*}

The following materials illustrate representative electronic-structure and chemical motifs covered by the benchmark.
\begin{itemize}
\item HgTe (mp-2730) is a narrow-gap semiconductor with band inversion and strong spin-orbit coupling.
\item SnTe (mp-1883) is a topological crystalline insulator.
\item TiCoSb (mp-5967) is a half-Heusler thermoelectric compound with three atoms in the primitive cell and space group 216.
\item PtS$_2$ (mp-762) is a layered transition-metal dichalcogenide for which semilocal DFT overestimates the band gap relative to experiment.
\item AgBr (mp-23231) is a representative case where semilocal DFT substantially underestimates the experimental band gap.
\item Cu$_2$O (mp-361) is a cuprite-structure d-electron system with space group 224.
\item HfS$_2$ (mp-985829) is a trigonal transition-metal dichalcogenide with space group 164.
\item YbTe (mp-1779) is an f-electron rocksalt compound.
\item NiO (mp-19009) is a Mott insulator whose ground state requires antiferromagnetic ordering.
\item SnO$_2$ (mp-856) is a rutile-structure oxide with space group 136 and a band gap that is strongly underestimated by semilocal DFT.
\item KTaO$_3$ (mp-3614) is a perovskite compound commonly regarded as a ferroelectric parent material.
\item CaF$_2$ (mp-2741) is an ionic wide-gap insulator with the fluorite structure.
\item BiClO (mp-22939) is a layered oxyhalide compound relevant to photocatalysis.
\end{itemize}

\section{Additional Analysis}
\label{app:additional-analysis}
\subsection{Recovery Analysis}
\label{app:recovery}

The closed-loop architecture routes failed or questionable steps through multiple recovery paths: LLM-driven parameter revision, rule-based fixes for known error signatures, and reflector-initiated plan modification. Table~\ref{tab:recovery} reports recovery statistics for AutoDFT-Full and AutoDFT-OpenLoop on VASPBench.

Under AutoDFT-Full, GPT-5.2 triggers recovery on 71 evaluated attempts, of which 31 are ultimately brought to convergence, yielding a recovery rate of 43.7\%. Claude Sonnet 4.6 and Gemini 3.1 Pro Preview require recovery less frequently (17 and 8 tasks, respectively), but they achieve higher recovery rates (52.9\% and 62.5\%). Plan modification through the step reflector is the most frequently triggered path, particularly for GPT-5.2, with 233 reflector-initiated modifications. This result confirms that iterative plan revision is a critical mechanism in the closed-loop architecture. In contrast, AutoDFT-OpenLoop lacks LLM-based recovery capabilities, which leads to a recovery rate of 0.0\%, highlighting the limitations of fixed recovery heuristics.

\begin{table*}[t]
\centering
\small
\caption{Recovery statistics on VASPBench. LLM: LLM-driven recovery attempts; Rule: rule-based recovery; Refl: reflector-initiated plan modifications; Need: attempts requiring recovery; Res: tasks ultimately resolved. Recovery statistics for baselines are reported only for GPT-5.2.}
\label{tab:recovery}
\begin{tabular}{llcccccc}
\toprule
Config & Backbone & LLM & Rule & Refl & Need & Res & Rate (\%) \\
\midrule
OpenLoop & GPT-5.2 & 0 & 12 & 0 & 6 & 0 & 0.0 \\
Rule-Based & GPT-5.2 & 0 & 23 & 0 & 12 & 1 & 8.3 \\
\midrule
\multirow{3}{*}{Full} & GPT-5.2 & 139 & 15 & 233 & 71 & 31 & 43.7 \\
& Claude Sonnet 4.6 & 33 & 0 & 23 & 17 & 9 & 52.9 \\
& Gemini 3.1 Pro Preview & 13 & 0 & 17 & 8 & 5 & 62.5 \\
\bottomrule
\end{tabular}
\end{table*}

\subsection{Execution Cost and Resource Efficiency}
\label{app:execution-cost}

\begin{table*}[t]
\centering
\small
\caption{Average LLM usage per task on VASPBench. Calls are broken down by agent stage: strategic planner (Plan), step planner (Step), reasoning monitor (Mon), recovery agent (Rec), and step reflector (Refl). OpenLoop uses a single planning call with rule-based execution.}
\label{tab:token-cost}
\begin{tabular}{llccccccc}
\toprule
Config & Backbone & Calls & Tokens & Plan & Step & Mon & Rec & Refl \\
\midrule
OpenLoop & GPT-5.2 & 1.0 & 3.9k & 1.0 & 0.0 & 0.0 & 0.0 & 0.0 \\
\midrule
\multirow{3}{*}{Full} & GPT-5.2 & 23.2 & 130k & 1.0 & 10.2 & 0.3 & 2.1 & 9.6 \\
& Claude Sonnet 4.6 & 9.8 & 59k & 1.0 & 3.9 & 0.2 & 0.9 & 3.7 \\
& Gemini 3.1 Pro Preview & 8.1 & 42k & 1.0 & 3.4 & 0.0 & 0.4 & 3.3 \\
\bottomrule
\end{tabular}
\end{table*}

\begin{table*}[t]
\centering
\small
\caption{Aggregate LLM-token and VASP execution cost on VASPBench. We report the number of tasks ($n$), successful tasks, total LLM tokens, total GPU-hours, GPU-hours per task, and GPU-hours per successful task. Rule-Based uses no LLM calls. OpenLoop performs a single planning call followed by rule-based execution, while Full performs closed-loop planning, monitoring, recovery, and reflection.}
\label{tab:execution-cost}
\begin{tabular}{llrrrrr}
\toprule
Backbone & Baseline & $n$ & Success & Tokens & GPU h & GPU h / task \\
\midrule
GPT-5.2 & Rule-Based & 34 & 23 & 0 & 1578.7 & 46.4 \\
GPT-5.2 & OpenLoop & 34 & 28 & 131.5k & 749.8 & 22.1 \\
GPT-5.2 & Full & 34 & 32 & 9.38M & 1,497.7 & 44.0 \\
\midrule
Claude Sonnet 4.6 & Rule-Based & 34 & 23 & 0 & 1,582.8 & 46.6 \\
Claude Sonnet 4.6 & OpenLoop & 34 & 26 & 125.6k & 1,265.4 & 37.2 \\
Claude Sonnet 4.6 & Full & 34 & 29 & 2.20M & 494.4 & 14.5 \\
\midrule
Gemini 3.1 Pro Preview & Rule-Based & 34 & 23 & 0 & 1,566.3 & 46.1 \\
Gemini 3.1 Pro Preview & OpenLoop & 34 & 26 & 147.7k & 1,369.4 & 40.3 \\
Gemini 3.1 Pro Preview & Full & 34 & 31 & 1.42M & 768.5 & 22.6 \\
\bottomrule
\end{tabular}
\end{table*}

A natural concern with closed-loop execution is that it requires additional LLM calls and therefore incurs a higher token cost. However, for VASP-based workflows, LLM usage is only one component of the overall cost: failed or poorly monitored calculations can consume many GPU-hours before reaching a hard timeout. We therefore analyze both LLM-token usage and VASP execution time.

Table~\ref{tab:token-cost} reports the per-task LLM usage by agent stage, and Table~\ref{tab:execution-cost} reports the aggregate token and VASP execution cost. As expected, AutoDFT-Full uses substantially more LLM tokens than OpenLoop, since it invokes the step planner, monitor, recovery agent, and reflector during execution. Nevertheless, the additional token cost remains small compared with the cost of the underlying VASP calculations. More importantly, closed-loop monitoring can reduce wasted VASP time by terminating unpromising runs early and triggering recovery before a calculation reaches its hard timeout.

This effect is especially pronounced for Claude Sonnet 4.6 and Gemini 3.1 Pro Preview. With Claude Sonnet 4.6, AutoDFT-Full reduces total VASP time from 1,582.8 GPU-hours under Rule-Based execution to 494.4 GPU-hours, while improving success from 23/34 to 29/34. With Gemini 3.1 Pro Preview, AutoDFT-Full similarly reduces total VASP time from 1,566.3 to 768.5 GPU-hours while improving success from 23/34 to 31/34. Thus, for these backbones, closed-loop execution improves both quality and resource efficiency: the extra LLM calls are offset by substantially fewer wasted VASP hours.

GPT-5.2 exhibits a different trade-off. AutoDFT-Full achieves the highest success rate, solving 32/34 tasks compared with 23/34 for Rule-Based and 28/34 for OpenLoop, but it also consumes more VASP time. This is likely because GPT-5.2 triggers more recovery iterations and retries, increasing the amount of downstream computation. Overall, these results show that AutoDFT exposes a tunable trade-off between LLM-token cost, VASP-hour cost, and final task success. In favorable regimes, closed-loop execution is not merely more accurate, but also more resource-efficient.

\subsection{Controlled Baselines and Repeated-Run Evaluation}
\label{app:controlled-ablations}

\paragraph{Baseline construction.}
To distinguish improvements from stronger upfront parameterization and those from execution-time adaptation, we evaluate four configurations on the same 34 VASPBench tasks using Gemini 3.1 Pro Preview. \textit{OpenLoop} generates a complete workflow in an initial LLM call and relies on heuristic parameter templates during execution. \textit{Static-Plan} instead generates both the complete workflow and all VASP input parameters, including INCAR and KPOINTS settings, in one upfront LLM call; the resulting plan and parameters are fixed before execution. \textit{Static-Plan + LLM Repair} adds LLM-driven recovery when VASP execution errors arise, while retaining the fixed initial plan and omitting history-conditioned per-step parameter generation, LLM-based monitoring, and post-convergence reflection. \textit{AutoDFT-Full} includes these closed-loop mechanisms as described in Section~\ref{sec:method}. These intermediate baselines separate upfront parameter generation from reactive error handling, although they do not isolate every component of the Full system individually.

\paragraph{Repeated-run results.}
We repeat each configuration three times on the full 34-task benchmark. Table~\ref{tab:app-repeated-runs} lists the number of successfully completed tasks in each run and reports the sample standard deviation of the three task-level success percentages. OpenLoop achieves 75.5\% $\pm$ 1.7\% across runs, while Static-Plan achieves 76.5\% $\pm$ 2.9\%. Thus, replacing heuristic templates with LLM-generated parameters fixed in advance does not materially improve mean task success in this setting. Adding reactive LLM Repair raises mean success to 81.4\% $\pm$ 4.5\%. AutoDFT-Full achieves the highest mean, 91.2\% $\pm$ 0.0\%, solving 31 of the 34 tasks in each of the three runs. The zero observed standard deviation refers to the aggregate success count across these three runs; it does not imply that internal execution trajectories or agent decisions are deterministic.

\begin{table*}[t]
\centering
\caption{Repeated-run task success on VASPBench using Gemini 3.1 Pro Preview. Each run evaluates all 34 tasks. Mean and sample SD are calculated from the success percentages across the three runs, not from independent per-task confidence intervals.}
\label{tab:app-repeated-runs}
\begin{tabular}{lcccc}
\toprule
Configuration & Run 1 & Run 2 & Run 3 & Mean $\pm$ sample SD \\
\midrule
OpenLoop & 26/34 & 25/34 & 26/34 & 75.5\% $\pm$ 1.7\% \\
Static-Plan & 25/34 & 27/34 & 26/34 & 76.5\% $\pm$ 2.9\% \\
Static-Plan + LLM Repair & 29/34 & 26/34 & 28/34 & 81.4\% $\pm$ 4.5\% \\
AutoDFT-Full & 31/34 & 31/34 & 31/34 & \textbf{91.2\% $\pm$ 0.0\%} \\
\bottomrule
\end{tabular}
\end{table*}

\paragraph{Compute and inference cost.}
Table~\ref{tab:app-ablation-cost} provides approximate per-task resource measurements for the configurations considered in this ablation. Static-Plan requires fewer VASP GPU-hours per task than AutoDFT-Full (12.4 versus 22.6), so higher task success should not be interpreted as universally reducing compute. However, Static-Plan + LLM Repair and Full require nearly identical VASP compute (22.2 versus 22.6 H100 GPU-hours/task) while achieving different mean success rates (81.4\% versus 91.2\%). The latter comparison suggests that the remaining gain is not simply explained by a large difference in VASP execution time. Full uses substantially more LLM tokens, approximately 42k rather than 9k per task. At the Gemini pricing assumption of \$12 per million tokens used in Table~\ref{tab:main-results}, this difference amounts to approximately \$0.40 per task, compared with approximately \$45 for 22.6 H100 GPU-hours at \$2 per GPU-hour. These cost figures are approximate configuration-level summaries and should not be treated as per-run uncertainty estimates.

\begin{table*}[t]
\centering
\caption{Approximate per-task resource comparison for Gemini 3.1 Pro Preview ablations. Token counts are rounded; GPU-hours refer to VASP execution. The figures summarize the reported configurations and are not separate measurements for each repeated run.}
\label{tab:app-ablation-cost}
\begin{tabular}{lcc}
\toprule
Configuration & LLM tokens/task (approx.) & VASP H100 GPU-h/task \\
\midrule
OpenLoop & 4k & -- \\
Static-Plan & 5k & 12.4 \\
Static-Plan + LLM Repair & 9k & 22.2 \\
AutoDFT-Full & 42k & 22.6 \\
\bottomrule
\end{tabular}
\end{table*}

\paragraph{Interpretation.}
The controlled comparison supports two narrower conclusions. First, improving only the upfront parameter-generation mechanism leaves performance close to the existing OpenLoop baseline. Second, permitting an LLM to repair explicit errors increases mean task success but does not match the combined closed-loop system. Since Full jointly enables history-conditioned step planning, selective LLM monitoring, recovery, and reflection, the remaining Full-versus-Repair difference cannot be attributed to any single module without further ablation. The 34-task benchmark and three repeated runs also limit statistical generalization; we interpret the results as evidence within the evaluated task set rather than a precise estimate of performance over all DFT workloads.

\subsection{Step Reflector Ablation and Expert Inspection}
\label{app:reflector-evaluation}

\paragraph{Component ablation.}
We evaluate the contribution of the Step Reflector by removing it from AutoDFT-Full while otherwise retaining the Gemini 3.1 Pro Preview configuration. The complete system succeeds on 31/34 VASPBench tasks (91.2\%), whereas the version without the Step Reflector succeeds on 28/34 tasks (82.4\%). This 8.8-percentage-point difference provides direct end-to-end evidence that post-convergence assessment and its associated decisions contribute to task success in the tested setting. The ablation does not establish the contribution of each other closed-loop component, and it is not accompanied by repeated-run uncertainty estimates.

\begin{table}[t]
\centering
\caption{Step Reflector ablation on VASPBench with Gemini 3.1 Pro Preview (34 tasks).}
\label{tab:app-reflector-ablation}
\begin{tabular}{lcc}
\toprule
Configuration & Success & Rate \\
\midrule
AutoDFT-Full w/o Step Reflector & 28/34 & 82.4\% \\
AutoDFT-Full & 31/34 & \textbf{91.2\%} \\
\bottomrule
\end{tabular}
\end{table}

\paragraph{Qualitative scientific inspection.}
We additionally sampled 20 Step Reflector outputs during system testing and asked co-authors with DFT expertise to inspect the proposed decisions and reasoning in the context of the corresponding calculation outcomes and task objectives. No clear scientific errors were identified in these inspected outputs. However, some judgments were scientifically defensible without reflecting the level of nuance or experience expected from a senior DFT practitioner. This inspection is qualitative and does not supply a calibrated agreement score or a statistically representative estimate of reflector accuracy.

\paragraph{Limits of action-level evaluation.}
Reflector decisions often do not admit a unique correct label: accepting a result, tightening numerical settings, or revising a downstream step may all be reasonable under different accuracy and resource priorities. Consequently, a binary correct/incorrect assessment would not capture the distinction between scientific plausibility, efficiency, and expert-level judgment. Our evidence therefore supports the practical contribution of the Step Reflector at the workflow level, but does not establish that individual decisions are optimal or equivalent to expert decisions. A larger, structured, independently assessed evaluation remains an important direction for future work.

\subsection{Matched-Subset Materials Project Evaluation}
\label{app:matched-subset}

The coverage-aware Materials Project results in Table~\ref{tab:mp-results} compute mean absolute errors (MAEs) over the successfully completed predictions for each backbone and property. Because the numbers of completed predictions differ, these MAEs can reflect both property-prediction differences and changes in which materials are included. To control for evaluation coverage, we recompute all three property MAEs on the common subset of 16 materials for which GPT-5.2, Claude Sonnet 4.6, and Gemini 3.1 Pro Preview all produce the three requested properties.

\begin{table*}[t]
\centering
\caption{Matched-subset property MAEs on the same 16 Materials Project materials for all three backbones. All columns use the same material set; lower is better. Reference values are from Materials Project PBE calculations.}
\label{tab:app-matched-subset}
\begin{tabular}{lccc}
\toprule
Property & GPT-5.2 & Claude Sonnet 4.6 & Gemini 3.1 Pro Preview \\
\midrule
Bandgap (eV) & 0.80 & \textbf{0.30} & 0.46 \\
Magnetic moment ($\mu_B$) & 0.17 & \textbf{0.14} & 0.17 \\
Formation energy (eV/atom) & \textbf{0.29} & 0.36 & 0.44 \\
\bottomrule
\end{tabular}
\end{table*}

Table~\ref{tab:app-matched-subset} shows that magnetic-moment MAEs are close across backbones, ranging from 0.14 to 0.17~$\mu_B$. In contrast, bandgap and formation-energy errors continue to vary on the same materials. This matched-subset analysis removes the difference in evaluation coverage as an explanation for those remaining variations, but it does not determine their underlying causes. We retain the original coverage-aware results because completion rates and prediction accuracy answer complementary questions: the former measures how often the system produces a valid result, and the latter assesses agreement with reference values conditional on completion. Neither comparison removes the known approximations of the underlying DFT treatment.

\subsection{Preliminary Portability to Quantum ESPRESSO}
\label{app:qe-portability}

\paragraph{Backend adaptation.}
To assess whether AutoDFT's high-level workflow design can extend beyond VASP, we implemented a preliminary backend for Quantum ESPRESSO (QE). The high-level planning framework, accumulated execution history, and reflection logic are retained, while code-specific input generation, execution interfaces, and output parsing are adapted to QE. In particular, the backend generates the relevant \texttt{pw.x} control, system, and electrons settings, pseudopotential specifications, and \texttt{K\_POINTS} information instead of VASP-specific input files. It also parses the corresponding QE outputs into the information needed for subsequent workflow decisions. This separation between high-level reasoning and code-specific execution is consistent with the modular architecture described in Section~\ref{sec:method}.

\paragraph{Preliminary evaluation.}
Using Gemini 3.1 Pro, we evaluated the QE backend on five relatively simple tasks, and all five completed successfully. This provides an initial execution-based demonstration that the framework is not restricted to VASP input and output formats. The experiment is deliberately limited in scope: it does not establish performance on complex QE workflows, equivalence between VASP and QE numerical results, or robustness across a broader range of materials and calculation settings. A larger cross-code benchmark with explicit property-accuracy comparisons remains future work.

\subsection{Preliminary Evaluation with a Stronger Backbone}
\label{app:stronger-backbone}

A potential alternative explanation for the Full-versus-OpenLoop improvement is that sufficiently capable language models might produce a complete and reliable workflow before execution. We therefore conduct an additional preliminary comparison using GPT-5.6 Sol on the same 34 VASPBench tasks. Owing to the limited evaluation window, each GPT-5.6 Sol configuration is evaluated once. We report the original GPT-5.2 results for context in Table~\ref{tab:app-stronger-backbone}.

\begin{table}[t]
\centering
\caption{Single-run VASPBench task success for OpenLoop and Full with GPT-5.2 and GPT-5.6 Sol (34 tasks). GPT-5.6 Sol results are preliminary.}
\label{tab:app-stronger-backbone}
\begin{tabular}{lcc}
\toprule
Backbone & OpenLoop & Full \\
\midrule
GPT-5.2 & 28/34 & 32/34 \\
GPT-5.6 Sol & 28/34 & 31/34 \\
\bottomrule
\end{tabular}
\end{table}

With GPT-5.6 Sol, OpenLoop completes 28/34 tasks and Full completes 31/34 tasks, preserving a three-task difference in this single-run comparison. The result does not establish a statistically significant gap for GPT-5.6 Sol or a monotonic relationship between model capability and workflow success. It is, however, consistent with the architectural motivation for AutoDFT: downstream choices in DFT workflows can depend on execution-time information, including relaxed structures, convergence behavior, magnetic states, and electronic properties. Stronger initial reasoning may improve an upfront plan, but an entirely static plan cannot directly condition its later parameter choices on outcomes that have not yet been computed. The present evidence motivates, rather than conclusively establishes, the continued utility of history-conditioned execution with stronger backbones.

\subsection{Benchmark Scope and Generalization Boundaries}
\label{app:benchmark-scope}

VASPBench contains 34 tasks drawn from official VASP documentation and spans nine calculation types. Each task provides a target structure and natural-language objective, rather than a worked calculation example or predetermined INCAR, KPOINTS, or workflow specification. AutoDFT must therefore construct and execute the requested calculation from the available task information. This design provides coverage across multiple VASP functionalities, but it does not establish that the underlying documentation was absent from LLM pretraining data. We do not interpret VASPBench performance as a test on guaranteed unseen documentation or workflows.

The task distribution is intentionally not class-balanced. Static self-consistent calculations and structural relaxation occur frequently as prerequisites of more specialized workflows, whereas some spectroscopy and response calculations are less common in the curated task set. Some tasks involve more than one calculation type, so the per-type counts need not sum to 34. Categories with only one or two tasks cannot support precise conclusions about category-level success rates, even when their observed rates are 0\% or 100\%. The per-type results in Figure~\ref{fig:per-type} should therefore be viewed as a diagnostic breakdown rather than a population-level ranking of calculation-type difficulty.

The separate 20-material Materials Project evaluation broadens the range of tested material systems and checks numerical agreement for selected properties. The preliminary QE results in Appendix~\ref{app:qe-portability} provide an additional, limited cross-code check. Nevertheless, broader evaluation of unseen workflow families, more materials, alternative exchange-correlation functionals, and complex calculations on other electronic-structure packages is necessary before making stronger generalization claims.

\section{Qualitative Examples and Limitations}
\label{app:examples}
\subsection{Example LLM-Generated Skeletal Plan}\label{app:task12-skeletal-plan}

This section reproduces a representative LLM-generated \texttt{SkeletalPlan} for the bandgap calculation of NiO (mp-19009, GPT-5.2). The plan encompasses five steps targeting the bandgap of an antiferromagnetic Mott insulator. Fields are reproduced verbatim from the saved output.

\begin{lstlisting}[basicstyle=\ttfamily\footnotesize,breaklines=true,columns=fullflexible]
{
  "task_description": "Compute the electronic band gap of bulk NiO (mp-19009) from VASP calculations, using a physically appropriate magnetic/correlation treatment for this Mott/charge-transfer insulator.",
  "material_formula": "NiO",
  "material_type": "3D bulk",
  "overall_strategy": "Because NiO is strongly correlated and antiferromagnetic, compute the gap using spin-polarized DFT+U in the AFM-II ordering (requires a cell larger than the provided 2-atom primitive). Enforce AFM order by using an AFM-II-capable magnetic cell (>=4 atoms) and robust symmetry-safe settings (ISYM=0 plus tight SYMPREC). First construct/verify an AFM-II-capable magnetic cell with a from-scratch static SCF; then relax the AFM structure under DFT+U; then run a short static AFM-stabilization SCF; finally a high-quality dense-k static calculation to extract the gap from the eigenvalue spectrum. Add a band-structure line-mode run only if k-point locations / direct-vs-indirect character are needed.",
  "steps": [
    {
      "step_name": "Construct AFM-II cell + magnetism/symmetry sanity check",
      "objective": "Ensure the calculation cell and input settings can represent AFM-II NiO and will not be symmetrized back to Fm-3m or collapse to a nonmagnetic solution.",
      "computation_type": "Pre-processing + short VASP single-point (spin-polarized DFT+U) sanity check",
      "depends_on": [],
      "constraints": [
        "Build an AFM-II-capable cell (at least 4 atoms). Recommended: the 4-atom rhombohedral AFM-II primitive of rocksalt with AFM stacking along [111], or an 8-atom conventional cubic cell with alternating (111) Ni planes. Do not use the 2-atom primitive cell.",
        "Assign initial MAGMOM with opposite signs on the two Ni sublattices (e.g., +2 and -2 on Ni; ~0 on O). Target ~0 net moment.",
        "Force a from-scratch spin-polarized start: ISTART=0 and ICHARG=2.",
        "Disable symmetry explicitly: ISYM=0 with SYMPREC=1e-8.",
        "Use intended DFT+U on Ni 3d (same U as later steps).",
        "NSW=0 (no ionic updates); insulator-appropriate smearing (ISMEAR=0 small SIGMA, or ISMEAR=-5)."
      ],
      "success_criteria": [
        "Cell contains distinct Ni sublattices with opposite initial moments.",
        "SCF converges to an AFM-like state: nonzero local Ni moments with opposite signs and ~0 net cell magnetization.",
        "No restoration to Fm-3m / cubic 2-atom cell."
      ],
      "priority": "high",
      "condition": "Always run first; repeat if any later step reports Fm-3m symmetry or mag ~ 0.0.",
      "predefined_parameters": null,
      "metadata": {}
    },
    {
      "step_name": "AFM-II DFT+U variable-cell relaxation (symmetry-safe)",
      "objective": "Obtain a relaxed AFM-II ground-state geometry and magnetic configuration for NiO that does not collapse to a higher-symmetry/net-moment solution.",
      "computation_type": "VASP ionic + cell relaxation (spin-polarized DFT+U)",
      "depends_on": ["Construct AFM-II cell + magnetism/symmetry sanity check"],
      "constraints": [
        "Start from verified AFM-II cell.",
        "Initialize opposite Ni moments; target ~0 net magnetization.",
        "DFT+U on Ni 3d (consistent U).",
        "ISYM=0; tight SYMPREC.",
        "Ionic + cell DOFs allowed (variable-cell relaxation).",
        "Insulator-appropriate occupation/integration.",
        "Reject convergence if mag ~ 0.0 or net FM moment appears."
      ],
      "success_criteria": [
        "Electronic SCF stable.",
        "AFM solution preserved: opposite local Ni moments, ~0 net.",
        "Forces/stress within tolerances.",
        "Insulating or near-insulating electronic structure."
      ],
      "priority": "high",
      "condition": "Always run after AFM-II cell verification.",
      "predefined_parameters": null,
      "metadata": {}
    },
    {
      "step_name": "AFM insulating solution stabilization (static SCF, symmetry off)",
      "objective": "Lock in the intended AFM insulating charge/magnetic state on the relaxed structure so the dense-k run does not revert to a metallic solution.",
      "computation_type": "VASP static single-point SCF (spin-polarized DFT+U), moderate k-mesh",
      "depends_on": ["AFM-II DFT+U variable-cell relaxation (symmetry-safe)"],
      "constraints": [
        "Reuse relaxed structure + same XC+U.",
        "ISYM=0; tight SYMPREC; AFM moments consistent with relaxed sublattices.",
        "Insulator-appropriate integration.",
        "Save CHGCAR/WAVECAR for restart."
      ],
      "success_criteria": ["Converged AFM SCF (~0 net cell moment)", "Nonzero gap or insulating occupation."],
      "priority": "high",
      "condition": "Run after relaxation.",
      "predefined_parameters": null,
      "metadata": {}
    },
    {
      "step_name": "Dense-k static SCF on stabilized AFM structure (gap extraction)",
      "objective": "Compute well-converged eigenvalues over the BZ and extract the fundamental band gap (VBM/CBM) for the stabilized AFM NiO structure.",
      "computation_type": "VASP static SCF (DFT+U) with dense k-mesh",
      "depends_on": ["AFM insulating solution stabilization (static SCF, symmetry off)"],
      "constraints": [
        "Use stabilized AFM structure + same XC+U.",
        "ISYM=0; tight SYMPREC.",
        "Increase k-density vs Steps 1-2.",
        "Restart from previous CHGCAR/WAVECAR."
      ],
      "success_criteria": [
        "AFM preserved and converged.",
        "Non-zero fundamental gap extractable from eigenvalues.",
        "Gap stable under minor k-mesh refinement."
      ],
      "priority": "high",
      "condition": "Primary deliverable step.",
      "predefined_parameters": null,
      "metadata": {}
    },
    {
      "step_name": "Optional band-structure line-mode (direct/indirect confirmation)",
      "objective": "Resolve k-point locations of VBM/CBM and confirm direct vs indirect nature.",
      "computation_type": "VASP NSCF band structure (line-mode) using CHGCAR from previous step",
      "depends_on": ["Dense-k static SCF on stabilized AFM structure (gap extraction)"],
      "constraints": [
        "Reuse converged CHGCAR; do not re-relax.",
        "Use high-symmetry path matching the magnetic cell BZ.",
        "Same DFT+U + magnetic settings."
      ],
      "success_criteria": ["Clean NSCF bands along path", "VBM/CBM identified."],
      "priority": "medium",
      "condition": "Only run if direct/indirect character or k-extrema are required.",
      "predefined_parameters": null,
      "metadata": {}
    }
  ],
  "global_considerations": {
    "magnetism": "NiO requires AFM-II along [111]; the 2-atom cell cannot host AFM-II.",
    "correlation_method": "DFT+U on Ni 3d (Dudarev-style Ueff). Plain GGA/PBE often gives metallic / too-small-gap state.",
    "symmetry": "AFM lowers symmetry; safest to disable ISYM (and tighten SYMPREC) so it is not effectively cubic.",
    "bandgap_extraction": "Dense-k static SCF is primary; line-mode is optional for k-locations.",
    "spin_convergence": "Initial moments + symmetry-off + tight electronic convergence; verify site-resolved moments, not just total."
  },
  "reasoning_trace": "NiO is a prototypical correlated AFM insulator; a 2-atom cell cannot host AFM-II, and semilocal DFT often fails to open the gap. (0) AFM-II cell sanity-check; (1) relax with DFT+U; (2) stabilize the AFM insulating solution; (3) dense-k static SCF for the gap; (4) optional line-mode bands.",
  "material_id": "mp-19009",
  "metadata": {}
}
\end{lstlisting}

\subsection{Case Study: Mott insulator bandgap recovery on NiO}
\label{sec:case-studies}

Here, we present a qualitative case study to illustrate the reasoning behavior that distinguishes closed-loop execution from open-loop planning. NiO (mp-19009) is a prototypical Mott insulator with antiferromagnetic type-II (AFM-II) ordering and an experimental bandgap of 4.0~eV. Computing its bandgap correctly requires (i) constructing a cell that can represent the AFM-II magnetic order, (ii) initializing the MAGMOM with alternating signs on the Ni sites, (iii) applying a Hubbard $U$ correction, and (iv) suppressing symmetry operations that would collapse the AFM state back to the high-symmetry ferromagnetic or non-magnetic solution.

AutoDFT-Full begins with a strategic plan of three steps: AFM-II DFT+U relaxation, dense-$k$ static SCF, and gap extraction. The step planner generates appropriate parameters (ISPIN=2, MAGMOM = 2.0 $-$2.0 0.0 0.0, LDAUU = 6.2, ISYM = 0), and the first relaxation converges. However, the step reflector detects that the converged solution yields a 0.0~eV bandgap, inconsistent with the expected insulating behavior for this material class. The step reflector issues a \texttt{redo} decision.

Over the next several cycles, the system iterates through increasingly targeted strategies: it adjusts the Hubbard $U$ value, enforces \texttt{NUPDOWN = 0} to constrain the net magnetization, tightens the mixing parameters (\texttt{AMIX}, \texttt{BMIX\_MAG}) to stabilize the AFM solution, and inserts a dedicated ``insulating solution stabilization'' step that uses \texttt{ICHARG = 2} to force a fresh charge density initialization. After several plan modifications and one LLM-driven recovery, the system converges to an AFM-II insulating solution through a seven-step workflow spanning 34 reasoning records. The final calculation produces a non-zero bandgap consistent with the DFT+U reference for NiO.

An open-loop system would have reported the initial zero-bandgap result as a converged success, since all VASP output files are formally valid and no runtime error occurred. The physically incorrect result is only detectable through the kind of post-step scientific plausibility check that the step reflector provides.

\end{document}